\documentclass[12pt,a4paper]{article}

\usepackage{enumerate}
\usepackage{amsmath}
\usepackage{amsfonts}
\usepackage{amssymb}       %

\usepackage{hyperref}
\usepackage{color}

\usepackage{color}
\usepackage{graphicx}%
\usepackage{ifthen}

\usepackage{hyperref}

\newtheorem{Theorem}{Theorem}[part]
\newtheorem{Definition}{Definition}[part]
\newtheorem{Proposition}{Proposition}[part]

\newtheorem{Lemma}{Lemma}[part]
\newtheorem{Corollary}{Corollary}[part]
\newtheorem{Remark}{Remark}[part]

\newtheorem{Convention}{Convention}[part]
\makeatletter \@addtoreset{equation}{section}

\@addtoreset{Definition}{section}

\@addtoreset{Theorem}{section}

\@addtoreset{Proposition}{section}

\@addtoreset{Assumption}{section}

\@addtoreset{Corollary}{section}

\@addtoreset{Lemma}{section}

\@addtoreset{Remark}{section}

\@addtoreset{Example}{section}

\@addtoreset{Condition}{section}

\@addtoreset{Convention}{section}

\makeatother \makeatletter

\def \be{\begin{eqnarray}}
\def \ee{\end{eqnarray}}
\def \b*{\begin{eqnarray*}}
\def \e*{\end{eqnarray*}}

\def \E{\mathbb{E}}
\def \F{\mathbb{F}}

\def \N{\mathbb{N}}
\def \P{\mathbb{P}}
\def \Q{\mathbb{Q}}
\def \R{\mathbb{R}}

\def \[{[\,\!\![}
\def \]{]\,\!\!]}

\def \1{{\bf 1}}
\def \esssup{{\rm esssup}}

\def \proof{{\noindent \bf Proof}\quad}
\def \ep{\hbox{ }\hfill$\Box$}

\def \interior{{\rm int}}

\def\reff#1{{\rm(\ref{#1})}}

\def\Ac{{\cal A}}
\def\Bc{{\cal B}}

\def\Ec{{\cal E}}
\def\Fc{{\cal F}}
\def\Gc{{\cal G}}

\def\Lc{{\cal L}}
\def\Mc{{\cal M}}

\def\Lc{{\cal L}}
\def\Oc{{\cal O}}

\def\Tc{{\cal T}}
\def\Uc{{\cal U}}

\def\Xc{{\cal X}}

 \def\vs#1{\vspace{#1mm}}

\def\esssup{{\rm ess}\!\sup\limits}
\def\Cb{{\mathbf C}}

\def \E{\mathbb{E}}
\def \F{\mathbb{F}}

\def \N{\mathbb{N}}
\def \R{\mathbb{R}}

\def\P{\mathbb{P}}
\def\Q{\mathbb{Q}}

\def\T{\mathbb{T}}
\def\Ac{{\cal A}}
\def\Bc{{\cal B}}

\def\Ec{{\cal E}}
\def\Fc{{\cal F}}
\def\Gc{{\cal G}}

\def\Ic{{\cal I}}

\def\Lc{{\cal L}}
\def\Mc{{\cal M}}

\def\Oc{{\cal O}}

\def\Tc{{\cal T}}
\def\Uc{{\cal U}}

\def\Xc{{\cal X}}

\def\ep{\hbox{ }\hfill$\Box$}
\def\reff#1{{\rm(\ref{#1})}}
\def\be{\begin{eqnarray}}
\def\ee{\end{eqnarray}}
\def\beq{\begin{equation}}
\def\eeq{\end{equation}}
\def\b*{\begin{eqnarray*}}
\def\e*{\end{eqnarray*}}

\def\x{\times}

\def\={\;=\;}

\def\Pas{\mathbb{P}\text{-a.s.}}

\def\.{\;.}

\def\1{{\bf 1}}
\def\Esp#1{\mathbb{E}\left[#1\right]}

\def\Pro#1{\mathbb{P}\left[{#1}\right]}

\def\eqref#1{\reff{#1}}

\def\p{+}

\def\Kdeps{K^{\epsilon_{'}}}
\def\Kdepsb{K^{\bar \epsilon_{'}}}

\newcommand{\qed}{\ep}

\newcommand{\tdual}[1]{#1'} %
\def\bRp{\bar \R_{+}}
\def\bRpd{\bar \R_{+}^{2}}
\def\Mb{{\rm   M\hspace{-3,7mm}M}}
\def\Cb{{\rm C\hspace{-3mm}C}}

\def\normeM#1{   \|#1\|_{{\rm  M\hspace{-2,85mm}M}}}
\def\normeC#1{   \|#1\|_{{\rm  C\hspace{-2,05mm}C}}}
\def\Gr#1{{\rm Gr}(#1)}

\date{January 30, 2013}

\begin{document}

\title{Robust no-free lunch with vanishing risk, a continuum of assets and  proportional transaction costs}

\author{Bruno Bouchard\thanks{\small CEREMADE, Universit\'e Paris Dauphine and  CREST-ENSAE,  \texttt{bouchard@ceremade.dauphine.fr}. This author is supported by ANR grant Liquirisk.},     Emmanuel Lepinette\thanks{\small CEREMADE, Universit\'e Paris Dauphine, \texttt{lepinette@ceremade.dauphine.fr}}~~and 
 Erik Taflin\thanks{\small Chair in Mathematical Finance EISTI and AGM Universit\'e de Cergy,  \texttt{taflin@eisti.fr}}}

\maketitle

\begin{abstract} We propose a continuous time model for financial markets with proportional transactions costs and a continuum of risky assets. This is motivated by bond markets in which the continuum of assets corresponds to the continuum of possible maturities. Our framework is well adapted to the study of no-arbitrage properties and related hedging problems. In particular, we extend the Fundamental Theorem of Asset Pricing of Guasoni, R\'asonyi and L\'epinette (2012) which concentrates on the one dimensional case. Namely, we prove that the Robust No Free Lunch with Vanishing Risk assumption is equivalent to the existence of a Strictly Consistent Price System. Interestingly, the presence of transaction costs allows a natural definition of trading strategies and avoids all the technical and un-natural restrictions due to stochastic integration that appear in bond models without friction.  We restrict to the case where exchange rates are continuous in time and leave the general c\`adl\`ag case for further studies.   
\end{abstract}

\noindent {\bf Key words:}  No-arbitrage, transaction costs, continuous time bond market. 

\vs3

\noindent{\bf MSC 2010:}   91B25, 60G44.

\section{Introduction}
 
The main contribution of this paper is to construct a continuous time model for financial markets with proportional transaction costs allowing for a continuum of risky assets. Such a model should have two important properties: 1. financial strategies should be defined in a natural way ; 2. it should allow one to retrieve the main results already established in the ``finite dimensional price'' case.  Our model has both. 
 
Frictionless models with a continuum of assets have  already been proposed in the literature,  cf. \cite{Bj-Ka-Ru97}, \cite{Carmona-Tehr},  \cite{ET} and \cite{E.T Bond  Completeness}. However, working with infinite dimensional objects leads to important technical difficulties when it comes to  stochastic  integration.  This imposes non-natural restrictions on the set of admissible trading strategies, resulting in that even  markets with a {\sl unique  equivalent martingale measure} are incomplete, in the sense that the set of attainable bounded claims is generically only dense in $L^{\infty}$ and not closed. Other surprising pitfalls and counter-intuitive results were pointed out in   \cite{T09}. 

Introducing transaction costs allows one to reduce these problems. The main reason is that it naturally leads to a definition of wealth processes which does not require stochastic integration. Once frictions are introduced, one comes up with a more realistic but also more natural and somehow simpler model.

In \cite{BTa2010},  the authors studied for the first time an infinite dimensional setting within the family of models with proportional transaction costs. They considered a countable number of assets in a   discrete time framework, and imposed a version of the  \textit{efficient friction} condition, namely  that the duals of the solvency cones have non-empty interior. Since perfectly adapted to discrete time models, they  {studied} the  No-Arbitrage of Second Kind (NA2) condition, first introduced in  \cite{ras09} and \cite{Ras1}. They showed that it implies  the Fatou closure property of the set of super-hedgeable claims and noted that  this closure property is in general lost if the efficient friction condition is replaced by a weaker condition, such as only requiring the solvency cones to be proper (as in finite dimensional settings).

In \cite{BTa2010} also a dual equivalent characterization was given in terms of Many Strictly Consistent Price Systems (MSCPS condition), cf.  \cite{Kabanov1},  \cite{ras09}. These price systems are the counterpart of the martingale measures in frictionless markets, i.e.~the building  {blocks} of dual formulations for derivative pricing and portfolio management problems.

The main contribution of the present paper is to provide an extension of this model to a continuous time setting with a continuum of assets:  the price process is, roughly speaking (for details see (\ref{eq : Hyp  prix a t=0})--(\ref{eq : Hyp prix conti en temps})), a continuous process on a time interval $[0,T]$ with values in the space  $C([0,\infty])$ of continuous functions on $[0,\infty]$,   the assets being indexed by the elements in $[0,\infty]$. A portfolio process is then a process of bounded variations, taking values in the space of Radon measures $M([0,\infty])$ on $[0,\infty]$, i.e. the dual of  $C([0,\infty])$, when endowed with its sup norm. Taking into account the infinite dimension, we develop this into a Kabanov geometrical framework  {(cf. \cite{Kabanov1} for the finite dimensional case)}, with locally compact instantaneous solvency cones in $M([0,\infty])$  {endowed with its weak* topology}, their   dual cones being viewed as subsets of  $C([0,\infty])$. 

Within this model, we  study  the No-Free Lunch with Vanishing Risk property, which is admitted to be the natural no-arbitrage condition in continuous time frictionless markets  since the seminal paper of Delbaen and Schachermayer \cite{DelbSchac1}. As  \cite{GLR}, we consider a robust version  (hereafter RNFLVR),  {\sl robust} being understood in the sense  of  \cite{schach04}, see also  \cite{GRS}:  the no-arbitrage property should also hold for a model with slightly smaller transaction cost rates. It is now standard in the continuous time literature.

Within this framework, the  Fatou-closure (resp. weak*-closure) property of the set of super-hedgeable claims evaluated in num\'eraire (resp. in  num\'eraire units at $t=0$)  is established (Theorem \ref{thm: fermeture Xb}).  Moreover, by  {using Hahn-Banach} separation  and  measurable selection arguments, we prove   the existence of Strictly Consistent Price Systems, which turns out to be equivalent to the RNFLVR condition (Theorem \ref{thm: existence Z sous RNFLVR} and Theorem \ref{thm: intM non empty implies RNFLVR}). From these results, a super-hedging theorem would be easy to establish by following very standard arguments, compare for instance with \cite{B-C}, \cite{C-S} and \cite{DV-D-K}.

All these results are natural extensions of the finite dimensional case, which validates the well-posedness of our model.

Several subjects are left to future studies. First, we have chosen to consider continuous price and transaction costs processes. This restriction is motivated by our wish to separate the difficulties related to the infinite dimensional setting and the ones coming from possibly time discontinuous prices and exchange rates. The latter case would require an enlargement of the set of admissible strategies along the lines of \cite{C-S}.  We have no doubt that this is feasible within our setting and leave it to further studies. 
Second, the NA2 property of no-arbitrage (robust or not) could also be discussed in continuous time settings, see \cite{D2}. We also leave  this to further studies. 

\section{Model formulation} \label{S1}

We first briefly introduce some  notations that will be used throughout the paper. 

All random variables are supported by a filtered probability space $(\Omega,\Fc,$ $\F,\P)$, with $\F=(\Fc_{t})_{t\in \T}$ satisfying the usual conditions, $\T:=[0,T]$ for some $T>0$. Without loss of generality, we take $\Fc_{0}$ equal to $\{\Omega,\emptyset\}$  augmented with $\P$-null sets, and $\Fc_{T}=\Fc$. If nothing else is specified, assertions involving random variables or random sets are understood to hold modulo $\P$-null sets. We denote by $\Tc$ the set of all stopping times $\tau\in\T$.

As usually, for a sub $\sigma$-algebra $\Gc$ of $\Fc$ and a measurable space $(E,\Ec)$,  $L^{0}(\Gc;(E,\Ec))$ stands for the set (of equivalence classes modulo $\P$-null sets) of $\Gc/\Ec$-measurable $E$-valued random variables.  

For a topological space $E$, the Borel $\sigma$-algebra generated by $E$ is denoted $\Bc(E)$ and when no risk for confusion the terminology  ``measurable space $E$'' is used. For a sub $\sigma$-algebra  $\Gc$ of $\Fc$, this defines the notation $L^{0}(\Gc;E).$

For a normable (real) topological vector space $E$, we denote by  $L^{p}(\Gc;E)$, the linear subspace of elements $\zeta \in L^{0}(\Gc;E)$ such that, for a compatible norm $\|\cdot\|_{E}$, $\|\zeta\|_{E}$ has a finite moment   $\|\zeta\|_{L^{p}(\Gc;E)}$   of order $p$  if $p\in (0,\infty)$, and is essentially bounded if $p=\infty$. For  $p \geq 0$, $L^{p}(\Gc;E)$ is given its standard vector space topology.   For $E=\R$ or $\Gc=\Fc$, we sometimes omit these arguments.  %

For two topological spaces $E$ and $F,$ $C(E;F)$ is the set of continuous functions of $E$ into $F$. $C(E;\R)=C(E)$.

Let $E$ be a compact Hausdorff topological space  {(in the sequel all compact spaces are supposed to be  Hausdorff, if not stated differently).}  The Banach space $C_{\beta}(E)$ (resp. topological vector space $C_{\sigma}(E)$) is by definition the vector space  $C(E)$ endowed with its supremum norm $\|\cdot\|_{C(E)}$ (resp. with its weak $\sigma(C(E),M(E))$ topology), where  $M(E)$ is the vector space of real Radon measures on $E$, i.e. $M(E)$ is the topological dual of $C_{\beta}(E).$ Such Radon measures will always be identified with their unique extension to the completion of a regular Borel measure on $E$. We use the standard notation  $\mu(f)=\int_{E} f d\mu$ for $\mu \in  M(E)$ and all  $\mu$-integrable  real valued maps $f $ on $E$. If $f$ is  $\mu$-essentially bounded, we write  $f\mu$ to denote the measure in $M(E)$ defined by $(f\mu)(g)=\mu(fg)$ $\forall g \in C(E).$ The Banach space $M_{\beta}(E)$ (resp.  topological vector space $M_{\sigma}(E)$) is by definition the vector space  $M(E)$ endowed with its  total variation norm $\|\cdot\|_{M(E)}$ (resp with its weak* $\sigma(M(E),C(E))$ topology). The positive orthants of $C(E)$ and $M(E)$  are denoted by $C_{+}(E)$ and $M_{+}(E)$ respectively. We also use the notation $C_{>0}(E)$ for the set of continuous functions taking only strictly positive values.

If $\Gc$ is a sub $\sigma$-algebra of $\Fc$,  $G$ is a topological space and $F$ is a set-valued function  $\Omega \ni \omega \mapsto F(\omega) \subset G$, then $L^{0}(\Gc;F)$ is the subset of elements $f \in L^{0}(\Gc;G),$ such that $f(\omega) \in F(\omega)$ $\Pas$, so $L^{0}(\Gc;F)$ is the  set of  $\Gc/\Bc(G)$-measurable selectors of the graph   $\Gr{F}$ $:=$ $\{(\omega,e)\in \Omega\times G: e\in F(\omega)\}$.
In this context, we make the following convention concerning the topology of $G$: 
\begin{Convention} \label{conv: topology on K and K'}
When  $G = C(E)$ (resp. $G = M(E)$), by default $L^{0}(\Gc;F)$ is then the set of weakly (resp. weak*) measurable selectors of $\Gr{F}$.
\end{Convention}

When $E=\bRp:=\R_{+} \cup\{\infty\}$, the one point compactification of $\R_{+}$, we simply write $\Cb$ for $C(\bRp)$  {and} $\Mb$ for $M(\bRp)$. The objects $\Cb_\sigma,$ $\Cb_\beta$, $\Mb_\sigma$, $\Mb_\beta$, $\Cb_{+}$, $\Cb_{>0}$, $\Mb_{+}$, $\normeC{\cdot}$, $\normeM{\cdot}$,   $\Cb_{+\beta}$, etc. are defined in an obvious way with reference to $\Cb$ and $\Mb$.

Given a subset $Y \subset C(E)$, we say that a process $\zeta=(\zeta_{t})_{t\in \T}$ is $Y$-valued if $\zeta_{t}(\omega)\in Y$ for $(\omega,t) \in \Omega \times \T$ a.e.  $d\P \otimes dt.$ We say that it is strongly (resp. weakly) $\F$-adapted if $\Omega \ni \omega \mapsto \zeta_{t}(\omega)\in C(E)$ is $\Fc_{t}/\Bc(C_{\beta}(E))$-measurable (resp. $\Fc_{t}/\Bc(C_{\sigma}(E))$-measurable) for all $t\in\T$.  
The process $\zeta$ is said to be strongly continuous if $\zeta \in C(\T;C_{\beta}(E))$ $\Pas$

 Given a family of random Radon measures $\mu=(\mu_{t})_{t\in \T}$ on $E$ and $Y \subset M(E)$,  we say that $\mu$ is $Y$-valued if $\mu_{t}(\omega) \in Y$ for $(\omega,t) \in \Omega \times \T$ a.e.  $d\P \otimes dt.$ We say that it is weak* $\F$-adapted if the map $\Omega \ni \omega \mapsto \mu_{t}(\omega)$ is $\Fc_{t}/\Bc(M_{\sigma}(E))$-measurable for all $t\in \T$.

\subsection{Financial assets and transaction costs}

We first describe the financial assets. Since we want to allow for a continuum of assets, covering the case of bond markets, we model their evolution by a stochastic process with values in the set of curves on $\R_{+}$. More precisely, 
we consider a mapping 
$$
 \T\x \R_{+}\x \Omega \ni  (t,x,\omega)\mapsto S_{t}(x)(\omega):=S_{t}(x,\omega) \in (0,\infty),
$$
and interpret $S_{t}(x)$  as the value at time $t$ of the asset with index $x$.  

We make the following standing assumptions, throughout the paper: 
\be  \label{eq : Hyp  prix a t=0}
& S_0 \in C(\R_+) ~\text{is strictly positive and deterministic},&
\\
 \label{eq : Hyp  prix posi conti}
& S/S_0 \text{ is } \Cb_{>0} \text{-valued and  weakly $\F$-adapted},& 
\\ 
\label{eq : Hyp prix conti en temps}      
&  S/S_0   ~ \text{ is a strongly continuous process}.&
\ee

In models for   bond markets,   $x  \in  \R_{+}$ can be interpreted as the  maturity of a zero-coupon bond and  it is usually assumed  that  $x\mapsto S_t(x)(\omega)$ has (for a.e.~$\omega$) certain differentiability properties. In this paper,  we only impose its continuity and positivity. 
 Note that, although in applications to bond markets it is natural to model prices as a curve  $x\mapsto S(x)$ on $  \R_{+}$,
we here  assume that $\R_+ \ni x \mapsto S_t(x)(\omega)/S_0(x)$ has  an extension to $\Cb$. Similar conditions are satisfied in continuous time  models without transaction costs, cf.    \cite[Theorem 2.2]{ET}.

In this paper, we  consider a market with proportional transaction costs.  When transferring at time $t$ an amount $a(x,y)$ from the account invested in asset $x$ to the account invested in   asset $y$, the account invested in asset $y$ is increased by $a(x,y)$ and the account invested in asset $x$ is diminished by   $(1+\lambda_{t}(x,y))a(x,y)$. Otherwise stated 
 buying one unit of asset $y$ against	 units of asset $x$ at time $t$ costs $(S_t(y)/S_t(x))(1\p \lambda_t(x,y))$ units of asset $x.$
 
  The  mapping 
\be\label{eq: lambda dans [0,1]} 
\lambda~:~  \T\x \bRpd \x \Omega \ni  (t,x,y,\omega) \mapsto \lambda_{t}(x,y)(\omega) \in (0,\infty)
\ee
is assumed to have the following continuity and measurability  properties: 
\be  \label{eq : Hyp  lambda posi conti} 
& \lambda \text{ is } C(\bRpd) \text{-valued  and weakly $\F$-adapted},& 
\\ 
\label{eq : Hyp lambda conti en temps}      
&\lambda   ~ \text{ is a strongly continuous process},&
\\
\label{eq: inega triangulaire lambda}  &  1+\lambda_t(x,z) \leq (1+\lambda_t(x,y))(1+\lambda_t(y,z)) , \;\forall\; t \in \T, \, x,y,z\in \bar\R_+.&
\ee
The two first  assumptions are of technical nature. The ``triangular condition''  \reff{eq: inega triangulaire lambda} 
is natural from an economical point of view and does not limit the generality.  

The important assumption is contained in \reff{eq: lambda dans [0,1]} which imposes   (strictly) positive transaction costs on any exchange between two different assets.
This corresponds to a strong version of the usual {\sl efficient friction} assumption,  which was already imposed in continuous settings by \cite{GLR}, \cite{GRS} and \cite{KG}. See Remark \ref{rem: efficient frictions} below.

\begin{Remark}\label{rem : S et lambda adpate et continus}{\rm 
Since $\Cb_\beta$ is separable, the weak measurability  in (\ref{eq : Hyp  prix posi conti}) implies by Pettis' theorem  (cf. \cite[Sect. V.4]{Yosida}),  that $S/S_0$ is strongly $\F$-adapted. It follows from (\ref{eq : Hyp prix conti en temps}) that  $S /S_0 \in C(\T\x \bRp)$ $\Pas$ (cf. \cite[Ch. X, \S 1, nr. 4, Prop. 2 and nr. 6, Th. 2]{Bourb TG 5-10}). Similarly, $\lambda$ is strongly $\F$-adapted and $\lambda \in C(\T\x \bRpd)$   $\Pas$ }
\end{Remark}

\subsection{Wealth process} \label{sbsec: admissible strgs prtfs}

\subsubsection{Motivation through discrete strategies}

Before to provide a precise definition of the notion of {trading strategy} we shall use in this paper, let us consider the case of discrete in time and space strategies, in a deterministic setting. In such a context, we can model the money transfers from and to the accounts invested in assets $x_{i}\in \bar \R_{+}$, $i\ge 1$,  at times $s_{k}$, $k\ge 1$, by non negative real numbers $a_{s_{k}}(x_{j},x_{i})\ge 0$: the amount  of money transferred at time $s_{k}$    to the account invested in $x_{i}$ by selling some units of $x_j$. Since the price at time $s_{k}$ of the asset $x_{i}$ is $S_{s_{k}}(x_{i})$, the  net number of units of $x_{i}$ entering and exiting the portfolio at time $s_{k}$ is given by
\b*
  \frac{1}{S_{s_{k}}(x_{i})}\sum_{j\ge 1}\left[ a_{s_{k}}(x_{j},x_{i})-  (1+\lambda_{s_{k}}(x_{i},x_{j})) a_{s_{k}}(x_{i},x_{j})\right].
\e*
To obtain the time-$t$ value of these transfers, one needs to multiply by $S_{t}(x_{i})$: 
\b*
\frac{S_{t}(x_{i})}{S_{s_{k}}(x_{i})}\sum_{j\ge 1}\left[ a_{s_{k}}(x_{j},x_{i})-  (1+\lambda_{s_{k}}(x_{i},x_{j})) a_{s_{k}}(x_{i},x_{j})\right].
\e*
The global net value at time $t$ of all transfers to and from the account  invested in the asset $x_{i_{o}}$ on the time interval $[0,t]$ is then given by 
\b*
V_{t}(\{x_{i_{o}}\})=\sum_{j,k\ge 1}\1_{[0,t]}(s_{k}) \frac{S_{t}(x_{i_{o}})}{S_{s_{k}}(x_{i_{o}})}\left[ a_{s_{k}}(x_{j},x_{i_{o}})-  (1+\lambda_{s_{k}}(x_{i_{o}},x_{j})) a_{s_{k}}(x_{i_{o}},x_{j})\right].
\e*
These quantities will in general be random, but must be adapted in the sense that $a_{s_{k}}(x_{j},x_{i})$ is $\Fc_{s_{k}}$-measurable, for each $i,j,k\ge 1$. 

For   a real valued function $f$ on $\bRp$, let us set
\be\label{eq: deg G}
G_{t}(f)(s,x,y):=\1_{[0,t]}(s)\left[\frac{S_{t}(y)}{S_{s}(y)}  f(y) - \frac{S_{t}(x)}{S_{s}(x)} f(x) (1+\lambda_s(x,y)) \right]. 
\ee
Then,
$$
V_{t}(\{x_{i_{o}}\})=\int_{\T\x \bRpd}G_{t}(\1_{\{x_{i_{o}}\}})(s,x,y) dL(s,x,y)  
$$
where $L$ is the Borel measure on $\T\x \bRpd$ defined by 
$$
L(A\x B\x C):=\sum_{i,j,k\ge 1} a_{s_{k}}(x_{i},x_{j}) \delta_{s_{k}}(A) \delta_{x_{i}}(B)\delta_{x_{j}}(C)
$$
for $A\times B\x C $   in the Borel algebra   of $\T\x \bRpd$. 

If one wants to introduce an initial endowment $v=(v(\{x_{i}\}))_{i\ge 1}$ labeled in amount of money, then one has to convert it into time $t$-values so that the time $t$-value of the portfolio becomes 
$$
V_{t}(\{x_{i_{o}}\})=v(x_{i_{o}})S_{t}(x_{i_{o}})/S_{0}(x_{i_{o}})+\int_{\T\x \bRpd}G_{t}(\1_{\{x_{i_{o}}\}})(s,x,y) dL(s,x,y)  .
$$
Viewing $V_{t}$ and $v$ as a Radon measures on $\bRp$, this leads to  
\b*
V_{t}(f)= v(fS_{t}/S_{0})+L(G_{t}(f))\;,\;f \in \Cb.
\e* 

\subsubsection{Trading strategies and portfolio processes}

The discussion of the previous section shows that it is natural, in the presence of a continuum of assets, to model financial strategies and portfolio processes as measure-valued processes on  $\T\x \bRpd$ and $\bRp$ respectively. We now make this notion more precise. 

We recall that Radon measures are identified with their unique extension to  regular Borel measures.

\begin{Definition} \label{def: trans measure}
A trading strategy is a  $M_{+}(\T\x \bRpd)$-valued random variable $L$  such that the $M_{+}(\T\x \bRpd)$-valued process $ (L_{t})_{t\in \T}$ defined by 
\begin{equation} \label{eq: self-fin 1.1}
    L_{t}(f) =  L (f \1_{ [0,t]\x {\bar \R_+}^2 }),\; f\in C(\T\x \bRpd),\; t\in \T,
\end{equation}
is   weak*-adapted. We set by convention $L_{0-}\equiv 0$, and denote by $\Lc$ the collection of such processes.
\end{Definition}

Note that the above definition is a natural extension of the finite dimension case in which transfers are modeled by multidimensional c\`adl\`ag non-decreasing adapted processes.
  
We are now in position to  define the notion of portfolio processes.
 For 
 $f \in C(\T\x \bRp),$ we  set
\be\label{eq: def H}
H(f)(s,x,y):= f(s,y)-  f(s,x)(1+\lambda_s(x,y)),\; (s,x,y) \in \T\x\bRpd.
\ee
We note that $H$ is a linear continuous operator from $C_{\beta}(\T\x \bRp)$ to $ C_{\beta}(\T\x \bRpd)$ (and also when both spaces are endowed with the weak topology) and observe that according to the definition of $G$ in \reff{eq: deg G},  
\be\label{eq: lien G et H}
G_t(f)(s,x,y)= {1_{[0,t]}(s) H(\frac{1 \otimes (S_tf)}{S})(s,x,y),} \text{ for } f \in \Cb,
\ee
where for $g \in \Cb$ we define $1 \otimes g \in C( \T\x\bRpd)$ by $(1 \otimes g)(s,x)=g(x)$.

\begin{Definition}\label{def: trading strategy}  A portfolio process $V^{v,L}$  is a $\Mb$-valued process such that
\begin{equation} \label{eq: self-fin 4.1}
  V^{v,L}_t(f)=  v( fS_{t}/S_{0}) +  L(G_{t}(f)) \;,\;\mbox{ $ t\in \T$ and $ f\in \Cb$,}
\end{equation}
for some  trading strategy  $L$ and some initial endowment $v \in \Mb$. If $v= 0$, we simply write $V^{L}$. 
\end{Definition}
It follows   from     Proposition \ref{prop : joint measurablity}  {(b)} in the Appendix and from the continuity of $H$ that  $V^{v,L}$ is weak* $\F$-adapted.

\begin{Remark} \label{rmk: Lt adapted}
{
Two trading strategies $L, \tilde L \in \Lc$ give rise to the same portfolio process, i.e. $V^{L}=V^{\tilde L}$,  if and only if $(L-\tilde L) \circ H =0$. In fact,  $(L-\tilde L) \circ H =0$ if and only if $(L-\tilde L) \circ G_t =0$  for all $t \in \T$.
}

{
A related question is: If we only know that  $\tilde L$ is a  $M_{+}(\T\x \bRpd)$-valued random variable and that the portfolio process $V^{\tilde L}$, constructed as in (\ref{eq: self-fin 4.1}), is weak*-adapted, does it follow that $ \tilde L \in \Lc$,  {i.e.}  $t \mapsto  \tilde L |_{ [0,t]\x {\bar \R_+}^2 }$ is weak*-adapted ? The answer is no.  However, it follows from Corollary \ref{cor: measurable selection L} in the Appendix that there always exists  $L \in \Lc$ such that  $V^{L}=V^{\tilde L}$.
}
\end{Remark}

\subsection{Solvency cones and dual cones} \label{sec: olvency cones}
 
We first define, for $\omega \in \Omega$,   
\begin{equation} \label{eq: solv cone in  M(K) 1}
\tilde K (\omega) =\mathrm{cone}\{(1+\lambda_t(x,y)(\omega))  \delta_t\otimes \delta_x -  \delta_t \otimes \delta_y, \,  \delta_t \otimes \delta_x :    (t,x,y) \in (\T\x \bRpd)\cap \Q^{3}\},
\end{equation}
where $\mathrm{cone}$ denotes the convex cone (finitely) generated by a family.
The set $\tilde K_{t}(\omega):=\{\nu \in \Mb: ~\delta_{t}\otimes \nu \in \tilde K(\omega)\}$ coincides with solvent financial  positions at times $t \in \T \cap \Q$ in the assets $x \in \bRp\cap \Q$, i.e. portfolio values that can be turned into positive ones (i.e. elements of $\Mb_+$) by performing immediate transfers.  This corresponds to the notion of {\sl solvency cone} in the literature, see \cite{Kabanov1}.  
\vs2

We then  define $K(\omega)$ as the weak* closure in $M(\T\x \bRp)$ of $\tilde K(\omega)$.  Using  %
the a.s.~continuity of $(t,x,y)\mapsto \lambda_{t}(x,y)$ noted in Remark \ref{rem : S et lambda adpate et continus}, one easily checks that the  {(positive)}  dual cone $\tdual{K}(\omega)$ of  $K(\omega)$ in $M_{\sigma}(\T\x \bRp)$ is given by
\be 
  \tdual{K}(\omega) 
 &:=&\label{eq: dual cone in  C(K)}
 \{f \in  {C}(\T\x \bRp) :  \mu(f)\ge 0  \; \forall \mu \in K(\omega) \}
 \\
 &=& \{f \in C_{+}(\T\x \bRp) :  f(t,y) \leq (1+\lambda_t(x,y)(\omega)) f(t,x)\; \; \forall (t,x,y) \in \T\x \bRpd\}. \nonumber
\ee
{Given  $t \in \T$, the instantaneous solvency cone $K_t(\omega)$ in the state $\omega$ at time $t$}  and, what will be proved to be, their dual cones $\tdual{K}_t(\omega)$ are defined as 
\be 
\label{eq: inst solv cone-dual K} 
&K_t(\omega):=\{\nu \in \Mb: ~\delta_t\otimes \nu \in K(\omega)\},&
\\ \label{eq: inst solv cone-dual K'} 
 &\tdual{K}_t(\omega):={\rm cl}\{f(t,\cdot): \, f\in \tdual{K}(\omega) \},
\ee
in which   ${\rm cl}$ denotes the norm closure on  $\Cb$.

Before continuing  with our discussion, let us first state important properties of the above random sets. The proofs are provided at the end of this section.

For each $\omega\in \Omega$ and $t \in \T$, we denote by   $\interior{(\tdual{K}(\omega))}$ (resp. $\interior{(\tdual{K}_t(\omega))}$) the interior of  $\tdual{K}(\omega)$  (resp. $\tdual{K}_t(\omega)$) in  $C_\beta(\T\x \bRp)$ (resp. $\Cb_\beta$). Note that if the strong topology is replaced by the weak one, then the  interiors of $\tdual{K}(\omega)$  and $\tdual{K}_t(\omega)$ are always empty, since this is the case for  $C_+(\T \x \bRp)$ and $\Cb_+$.
The proofs of  the following results  are provided at the end of this section.

 \begin{Proposition}\label{prop: int  K'} Fix $t \in \T$. Then $\Pas$,  $\interior{(\tdual{K}(\omega))}$ and $\interior{(\tdual{K}_t(\omega))}$ are non-empty,
\begin{equation}\label{eq: int K'}
 \begin{split}
   &\interior{(\tdual{K}(\omega))}=
 \{f \in C_{>0}(\T\x \bRp) : \\ & f(t,y) < (1+\lambda_t(x,y)(\omega)) f(t,x), \; \forall (t,x,y) \in \T\x \bRpd\},
 \end{split}
\end{equation}
\begin{equation}\label{eq: int K't}
 \begin{split}
   \interior{(\tdual{K}_t(\omega))}&=
 \{f \in \Cb_{>0}  : \\ & f(y) < (1+\lambda_t(x,y)(\omega)) f(x), \; \forall (x,y) \in  \bRpd\}, 
 \end{split}
\end{equation}
and
\begin{equation}\label{eq:  K't}
 \begin{split}
   \tdual{K}_t(\omega)&=
 \{f \in \Cb_{+}  : \\ & f(y) \leq (1+\lambda_t(x,y)(\omega)) f(x), \; \forall (x,y) \in  \bRpd\}.
 \end{split}
\end{equation}
\end{Proposition}

\begin{Remark}\label{rem: efficient frictions}  The fact that the cones $\tdual{K}_{t}$ have non-empty interior is an immediate consequence of the condition   $\lambda_{t}(x,y)>0$   $\forall( x,y)$ contained in  \reff{eq: lambda dans [0,1]}. The condition $\interior{\tdual{K}_t}\ne \emptyset$ is usually referred to as the {\sl efficient friction assumption}. In finite dimensional settings (i.e.~if $\bRp$ is replaced by a finite set), it is equivalent to the fact that the $K_{t}$ are proper or that  $\lambda_{t}(x,y)+\lambda_{t}(y,x)>0$ for all $x\ne y$, see e.g.~\cite{Kabanov1}. This last equivalence does not hold anymore when the dimension is not finite, see \cite[Remark 6.1]{BTa2010}. 
\end{Remark}

\begin{Proposition}\label{prop: propriete K et K'} Fix  $\tau\in \Tc$. Then,
\begin{enumerate}[\rm (a)]
\item  $K_{\tau}$ is  $\Pas$ closed in $\Mb_\sigma$ and it is $\Pas$ the dual cone of $\tdual{K}_{\tau}$ in  $\Cb_\sigma$. Moreover, ${\rm Gr}(K_{\tau}) \in \Fc_{\tau}\otimes \Bc(\Mb_\sigma)$.  
\item  $\tdual{K}_{{\tau}}$ is  $\Pas$ closed in $\Cb_\sigma$ and it is $\Pas$ the dual cone of $ {K}_{\tau}$ in $\Mb_\sigma$. Moreover, ${\rm Gr}( \tdual{K}_{\tau})$ $\in$ $\Fc_{\tau}\otimes \Bc(\Cb_\sigma)$. 
\end{enumerate}
\end{Proposition}

We now define the associated notion of liquidation value at $t\in \T$, the highest value in asset $0$ which can be obtained from a position $\nu \in L^{0}(\Fc_{t}; \Mb_\sigma)$ { at $t$} by liquidating all other positions in $ (0,\infty]$:
\be  \label{LiquiValueDef}
\ell_{t}(\nu)(\omega):=\sup\{x\in \R:~\nu(\omega)-x \delta_0 \in K_{t}(\omega) \}.
\ee
Observe that  the duality between $K_{t}$ and $\tdual{K}_{t}$ implies 
\be \label{eq: def ell par dualite}
\ell_{t}(\nu)(\omega)=\inf\{\nu(f)(\omega):~f\in \tdual{K}_{t}(\omega) \mbox{ s.t. } f(0)=1  \}. 
\ee 

The function $\ell_{t}$ inherits the measurability properties of Proposition \ref{prop: propriete K et K'}, as will be proved below. 
\begin{Proposition}\label{prop: propriete ell} $\ell_{\tau}(\nu)\in L^{0}(\Fc_{\tau})$ for all   $\tau\in \Tc$ and $\nu \in L^{0}(\Fc_{\tau};\Mb_{\sigma})$. 
\end{Proposition}

\begin{Remark}\label{rem : ell} {\rm Fix $\tau \in \Tc$.  Note that $f\in L^{0}(\Fc_{\tau};\tdual{K}_\tau)$ with $f(0)>0$   implies that 
$$
f /f(0)\ge (1+\lambda_{\tau}(\cdot,0))^{-1}  \ge \iota_\tau :=\min\{(1+\lambda_{\tau}(x,0))^{-1},\;x\in \bRp\}\in L^{0}(\Fc_\tau),
$$
in which we use the continuity assumptions \reff{eq : Hyp  lambda posi conti} and \reff{eq: lambda dans [0,1]}.
It thus follows from \reff{eq: def ell par dualite} that 
$
  \ell_{\tau} (  \nu)\ge \iota_\tau \normeM{\nu}$    for all $\nu \in \Mb_+$. Note that $\min_{t\in \T} \iota_{t}>0$ $\Pas$ thanks to  {Remark \ref{rem : S et lambda adpate et continus} and \reff{eq: lambda dans [0,1]}}.
  }
\end{Remark}

{\bf Proof of Proposition \ref{prop: int  K'}.} Fix  $\omega\in \Omega$ such that $\lambda(\omega) \in C(\T\x \bRpd),$ which holds outside a set of measure zero according to Remark \ref{rem : S et lambda adpate et continus}.
 
Let $f \in \interior{(\tdual{K}(\omega))}$, i.e. for some $\epsilon >0$, $f+B(\epsilon) \subset  \tdual{K}(\omega)$, where $B(\epsilon)$ is the open ball in $C(\T\x \bRp)$ of radius $\epsilon$ centered at $0$. Since $\T\x \bRp$ is compact, it   follows from (\ref{eq: dual cone in  C(K)}) that such an $\epsilon$ exists if and only if formula (\ref{eq: int K'}) holds.

Let $e \in  C(\T\x \bRp)$ be the constant function taking the value $1$. Then  $e \in \interior{(\tdual{K}(\omega))}$ according to (\ref{eq: int K'}), since  $\lambda(\omega)$ has a strictly positive minimum on $\T\x \bRpd$ by compactness and continuity.

Let $A_t$ be the right hand side in the equality (\ref{eq: int K't}). $A_t$  is non-empty since it contains the positive  constant functions, recall \reff{eq: lambda dans [0,1]}.

We define the linear continuous operator $P_t : C_\beta(\T\x \bRp) \rightarrow \Cb_\beta$ by $(P_tf)(x)=f(t,x).$  Being also surjective,  $P_t$ is an open mapping. Therefore $\Oc_t:=P_t(\mathrm{int}(\tdual{K}(\omega)))$  is a non-empty open set.

For the moment, we make the hypothesis that
$$\Oc_t= A_t.$$
Since $\mathrm{int}(\tdual{K}(\omega))$ and $A_t$ are non-empty convex cones, their closures coincide with the closures of their interiors. The continuity of $P_t$ thus ensures that  $\tdual{K}_t(\omega)=\mathrm{cl}(\Oc_t) = \mathrm{cl}(A_{t})$. This proves equality (\ref{eq:  K't}). Taking the interior of both sides of this equality gives (\ref{eq: int K't}).

Finally, we prove the above hypothesis $\Oc_{t}=A_{t}$. The inclusion  $\Oc_t \subset A_t$ follows trivially, by  definition (\ref{eq: inst solv cone-dual K'}) and equality  (\ref{eq: int K'}).
To prove the inclusion $A_t \subset \Oc_t$,  fix $g \in A_t$ and a function $\phi \in C_+(\T)$   such that $0\le \phi \leq 1$ on $\T$, $\phi(t)=1$,  and $\mathrm{supp} (\phi) \subset [t-\delta,t+\delta] \cap \T$ for some $\delta >0$. Define $f \in C_+(\T\x \bRp)$ by $f(s,x)=\phi(s) g(x)+(1-\phi(s))$. Then $P_t(f)=\phi(t)g=g$. Since the unit constant function $e$ belongs to    $ \interior{(\tdual{K}(\omega))}$,  a  compactness and continuity argument allows to choose $\delta >0$ small enough such that  $f \in \interior{(\tdual{K}(\omega))}$ given by (\ref{eq: int K'}), recall Remark \ref{rem : S et lambda adpate et continus}.
\ep

\vs2

{\bf Proof of Proposition \ref{prop: propriete K et K'}.} 1. We fix $\omega \in \Omega$ and set $t:=\tau(\omega)$ to alleviate the notations. Let $M_{t}(\bRp)$ be the subspace of measures $\mu \in  M(\T \x \bRp)$  such that $\mathrm{supp}(\mu) \subset \{t\} \x \bRp$. Then  $\mu \in M_{t}(\bRp)$ if and only if $\mu = \delta_t\otimes  \nu$ with $\nu \in \Mb$ and $M_{t}(\bRp)$ is a closed subspace of $M_\sigma(\T \x \bRp)$.  Let     $M_{t\sigma}(\bRp)$ be $M_{t}(\bRp)$ endowed   with the induced topology, as a  subspace of $M_\sigma(\T \x \bRp)$.  Then $A: =K(\omega)  \cap M_{t}(\bRp)$ is a closed convex cone in $M_{t\sigma}(\bRp)$. The linear mapping $M_{t\sigma}(\bRp) \ni  \delta_t \otimes \nu \mapsto \nu \in \Mb_\sigma$ is a continuous linear isomorphism. Under this mapping, $K_t(\omega)$ is the image of $A$, so  $K_{t}(\omega) $ is a closed convex cone in $\Mb_\sigma$.
 
2. Here again, we  fix $\omega \in \Omega$ and set $t:=\tau(\omega)$ to alleviate the notations.  By definition, $\tdual{K}_{t}(\omega)$ is $\Pas$ a closed convex cone in $\Cb_\beta.$ Being convex, it is then also closed in $\Cb_\sigma.$  Clearly, formula (\ref{eq:  K't}) of Proposition \ref{prop: int  K'} shows that $\tdual{K}_{t}(\omega)$ is  the dual cone of $K_{t}(\omega)$ in $\Mb_\sigma$.
Since $K_{t}(\omega) $ is convex and closed in $\Mb_\sigma,$ it now follows by the bipolar theorem  that the  dual cone of $ \tdual{K}_{t}(\omega)$ in  $\Cb_\sigma$ is  $K_{t}(\omega)$. 
 
3. We now prove the measurability properties.  
a. We start with   $\tdual{K}_{\tau}$. For $f\in \Cb$ and $t\le \T$, let us set  
\be\label{eq: def F}
\hat F_{t}(f)(\omega):=\inf_{(x,y)\in \bRpd} F_{t,x,y}(f)(\omega),
\ee
where 
$$
F_{t,x,y}(f)(\omega):=\big(f(x)(1+\lambda_{t}(x,y)(\omega))-f(y))\big) \wedge f(x).
$$
Note that, for $f\in \Cb$, 
\b*
 (\omega, f)\in   {\rm Gr}( \tdual{K}_{\tau})^{c} 
 &\mbox{ if and only if }&
 \hat F_{\tau(\omega)}(f)(\omega)<0. 
\e*
For $n\ge 1$ and $0\le k\le 2^{n}$, set $s^{n}_{k}:=k2^{-n}t$, for some $t\in \T$, and let $(x_{l},y_{m})_{l,m\ge 1}$ be dense in $\bRpd$. Then, the above, combined with the continuity of $\lambda$ stated in Remark \ref{rem : S et lambda adpate et continus} and the compactness of $\T\x\bRpd$, implies that 
$$
A_{t}:={\rm Gr}( \tdual{K}_{\tau})^{c}\cap (\{\tau \le t\}\x  \Cb)=\cap_{N\ge 1 }\cup_{n\ge N} \cup_{k=1}^{2^{n}} \cup_{l,m\ge 1 }A_{t,n}^{k,l,m}
$$
where 
$$
A^{k,l,m}_{t,n}:=\{(\omega,f)\in \Omega\times \Cb: \tau(\omega)\in (s^{n}_{k-1},s^{n}_{k}]  \mbox{ and } F_{s^{n}_{k},x_{l},y_{m}}(f)(\omega)<0\}
$$
with the convention $(s^{n}_{0},s^{n}_{1}]=[0,s^{n}_{1}]$.    The mapping $(\omega, f)\mapsto  (\lambda_{s^{n}_{k}}(x_{l},y_{m})(\omega),$ $\delta_{y_m}(f),\delta_{x_{l}}(f))$  $=$ $(\lambda_{s^{n}_{k}}(x_{l},y_{m})(\omega),f(y_{m}),f(x_{l}))$ of $(\Omega \x \Cb_\sigma,\Fc_{t}\otimes \Bc(\Cb_{\sigma}))$ into $\R^3$ is a Carath\'{e}odory function, i.e. measurable with respect to $\omega$ and continuous  with respect to $f$, hence $\mathcal{F}_t\otimes $ $\Bc(\Cb_{\sigma})$-measurable. By continuous compositions, so is the mapping 
$(\omega, f)\mapsto  F_{s^{n}_{k},x_{l},y_{m}}(f)(\omega)$.   Hence, $A_{t}\in \mathcal{F}_t\otimes $ $\Bc(\Cb_\sigma)$. By arbitrariness of $t\in \T$, this shows that ${\rm Gr}( \tdual{K}_{\tau})$ $\in $ $\mathcal{F}_\tau\otimes $ $\Bc(\Cb_\sigma)$. For later use, note that minor modifications of  the above arguments   show  that
\be\label{eq: Ftau Fctau-meas}
   \hat F_{\tau}(f)\in L^{0}(\Fc_{\tau})\; \mbox{ for all }\; f\in \Cb.
\ee 
 
 b. It remains to discuss the measurability of ${\rm Gr}(K_{\tau})$. It will  follow from the $\Pas$ duality between $K_{\tau}$ and $\tdual{K}_{\tau}.$ We first note that  
\be\label{eq: cara K' par F}
\mbox{$g\in  {\rm int}( \tdual{K}_{\tau}(\omega))$  if and only if $g \in \Cb$ and $\hat F_{\tau}(g)(\omega)> 0$,}
\ee
 where $\hat F$ is defined as in \reff{eq: def F}. 
Let  $(f_{n})_{n\ge 1}$ be a dense family of  $\Cb_\beta$ and set 
$$
B_{n}:=\{(\omega,\nu)\in \Omega\times \Mb:  \max\{\nu(f_{n}),-\hat F_{\tau(\omega)}(f_{n})(\omega)\}\ge 0\},\; n\ge 1. 
$$
The assertion \reff{eq: Ftau Fctau-meas} implies   that $B:=\cap_{n} B_{n}$ is an element of  $\mathcal{F}_\tau\otimes \Bc(\Mb_\sigma)$.
To conclude the proof, we now show that ${\rm Gr}(K_{\tau})=B$. The inclusion ${\rm Gr}(K_{\tau})\subset B$ follows from  \reff{eq: cara K' par F}.  To obtain the converse inclusion, we first recall that ${\rm int}(\tdual{K}_{\tau(\omega)}(\omega))$ is a non-empty convex cone so that its norm closure in $\Cb_{\beta}$ coincides with $\tdual{K}_{\tau(\omega)}(\omega)$. 
This implies that $\nu \in K_{\tau(\omega)}(\omega)$ whenever  $\nu(g)\ge 0$ for all $g \in {\rm int}(\tdual{K}_{\tau(\omega)}(\omega))$, or, equivalently,  if $\nu(g)\ge 0$ for all $g \in \Cb$ such that $-\hat F_{{\tau(\omega)}}(g)(\omega)<0$, recall  \reff{eq: cara K' par F}. By a.s.~continuity of $g\in \Cb_{\beta}\mapsto \hat F_{\tau(\omega)}(g)(\omega)$, this is satisfied by any $(\omega,\nu) \in B$.
\ep
\\

{\bf Proof of Proposition \ref{prop: propriete ell}.} The result follows from Proposition \ref{prop: propriete K et K'} and the fact that, for $c_{o}\in \R$, 
$$
\{\omega\in \Omega: \ell_{\tau(\omega)}(\nu(\omega))(\omega)<c_{o}\}=\cup_{c\in \Q\cap(-\infty,c_{o}) } \{\omega\in \Omega: (\omega,\nu(\omega)-c\delta_{0})\in {\rm Gr}( {K}_{\tau})\}.
$$
\ep

\section{Robust no free lunch with vanishing risk and closure properties} \label{sec : NFLVR}

\subsection{Definitions}

We are now in position to define the notion of no-arbitrage we shall consider. As in \cite{GLR}, we use the robust version of the No Free Lunch with Vanishing Risk criteria.
 For this purpose, we  restrict to strategies that are bounded from below in the following sense. 
\begin{Definition}\label{def: t-lower bound}
For  $c \in \R_+$,  $L^0_{b}(c)$ is the subset of random variables $\zeta \in L^0(\Fc_{T}; \Mb_\sigma)$  bounded from below by $c$ in the sense that
\begin{equation} \label{eq: l.b. rv 1}
\zeta + \frac{S_T}{S_0}\eta \in L^0(\Fc_{T}; K_{T}) \;,\, \text{ for some } \eta  \in \Mb  \text{ with  }  \normeM{\eta} \le c.
\end{equation}
The set of all   $\Mb$-valued random variables  bounded from below is 
$$
L^{0}_{b}:=\bigcup_{c\in \R_{+}} L^{0}_{b}(c).
$$
A strategy $L\in \Lc$ is said to be  bounded from below, if there exists
$\eta \in \Mb $ such that
$$
V_{t}^{L}+\frac{S_t}{S_0} \eta \in L^0(\Fc_{t};  K_{t}) \;,\, \text{ for   all $t\in \T$.} 
$$
We denote by $\Lc_{b}$ the set of such strategies, they are said to be admissible. The set of admissible strategies, for which the terminal portfolio values  are $c$-bounded from below is denoted by
$
\Lc_{b}(c):=\{L\in \Lc_{b}:  V^{L}_{T} \in L^{0}_{b}(c)\}. 
$
\end{Definition}

The set of bounded from below random claims that can be super-hedged starting from a zero initial endowment and by following an admissible strategy  is 
$$
\Xc_{b}^{T}:=\cup_{c\ge 0} \Xc_{b}^{T}(c)
$$
where
$$
\Xc_{b}^{T}(c):=\{X\in L^{0}_{b}(c): V^{L}_{T}-X\in L^0(\Fc_{T}; K_{T})  \;\mbox{ for some } L \in \Lc_{b}\}. 
$$

\vs2

The no-free lunch with vanishing risk property (NFLVR) is defined in a usual way.
\def\NFLVR{{\rm (NFLVR)}}
\def\NFLVRe{{\rm (NFLVR)}$^{\epsilon}$}
\def\RNFLVR{{\rm (RNFLVR)}}

\begin{Definition}[NFLVR]\label{def : NFLVR}  We say that  \NFLVR~holds  if for each sequence $(X_n,c_{n})_{n\ge 1}\subset \Xc_{b}^T\x  \R_+$ :

\noindent $\lim_n  c_n=0$ and $X_{n}\in  \Xc_{b}^T(c_{n})$ for all $n\ge 1$ imply $\limsup_{n} \ell_{T}(X_n)\le 0$ $\Pas$
\end{Definition}

In order to define a robust version of the above, one needs to consider  models with  transaction costs
\be\label{eq: lambda epsilon}
\lambda^{\epsilon}:=\lambda-\epsilon
\ee
{\sl strictly smaller} than $\lambda$. We denote by $\Upsilon$ the set of $C_{>0}(\bRpd)$-valued adapted processes $\epsilon$ such that the left-hand side of \reff{eq: lambda epsilon} satisfies the conditions  (\ref{eq: lambda dans [0,1]})--(\ref{eq: inega triangulaire lambda}).
\begin{Remark}\label{rmk: ex lambda epsilon}
An  easy example of a $\lambda^{\epsilon}$ is obtained by fixing $k \in (0,1)$ and setting $\epsilon_t(x,y)=1+\lambda_t(x,y) - (1+\lambda_t(x,y))^k$, $\forall (t,x,y) \in \T \x \bRpd.$ It is straightforward to check that $\lambda^{\epsilon}$ satisfies (\ref{eq: lambda dans [0,1]})--(\ref{eq: inega triangulaire lambda}). Due to compactness, and continuity and  strict positivity of $\lambda$, we have $\inf_{(t,x,y)}\epsilon_t(x,y) \in L^0(\Fc; (0,\infty))$.
\end{Remark}

We define $G^{\epsilon}$, $K^{\epsilon}$, $ \Kdeps$, $\ell^{\epsilon}_{T}$, $\Xc_{b}^{T\epsilon}$,\ldots, and  \NFLVRe~ as above with $\lambda^{\epsilon}$ in place of $\lambda$, for $\epsilon \in \Upsilon$.

\begin{Definition}[RNFLVR]\label{def: RNFLVR}  We say that  \RNFLVR~holds  if  \NFLVRe~holds for some   $\epsilon\in \Upsilon$.
\end{Definition}

The above definition is similar to Definition 5.2 in \cite{GLR}, except that they use a notion of simple strategies.   
\subsection{Closure properties}

The main result of this section is a  Fatou-type closure property for the set of terminal values of super-hedgeable claims $\Xc_{b}^{T}$.   
\begin{Definition}  We say that $(\mu_n)_{n\ge 1}\subset L^{0}(\Fc; \Mb)$ is  Fatou-convergent with limit $\mu$ if 
$(\mu_n)_{n\ge 1}$ converges $\Pas$ to $\mu$ in $\Mb_\sigma$ and   $(\mu_n)_{n\ge 1}\subset L_{b}^0(c)$ for some $c\in \R_{+}$.

A subset $F$ of  $ L^{0}(\Fc; \Mb)$ is said to be Fatou-closed if any Fatou-convergent sequence has a limit in $F$. 
\end{Definition}
It will readily imply that the corresponding set 
\be\label{eq: def hat Xc}
\hat \Xc_{b}^{T}=  \{S_{0}/S_{T} X \mbox{ for some } X\in \Xc_{b}^{T}\} 
\ee 
of super-hedgeable claims  {labeled in terms of numeraire units at $t=0$} is weak*-closed.

\begin{Theorem}\label{thm: fermeture Xb}  Assume that  \RNFLVR~holds. Then,   the set 
$\mathcal{X}_b^T$ is  Fatou-closed. Moreover, $\hat \Xc_b^T\cap L^{\infty}(\Fc_T; \Mb)$ is $\sigma( L^{\infty}(\Fc_T; \Mb), L^{1}(\Fc_T; \Cb))$-closed.
\end{Theorem}
The proof of Theorem \ref{thm: fermeture Xb}  will be split in several parts.   We first establish two boundedness properties which follow from our \RNFLVR~assumption (compare with \cite[Lemma 5.4, Lemma 5.5]{GLR}). 
\begin{Lemma} \label{lem : NFLVR implique borne proba val liqui}   Let \NFLVRe~hold for some $\epsilon \in \Upsilon$,  and fix $c\in \R_{+}$. Then,  the set $\ell_{T}^{\epsilon}(\Xc_{b}^{T\epsilon}(c)) \subset L^0(\Fc)$ %
is bounded in probability.   
\end{Lemma}

\proof If the assertion of the lemma is not true, then one can find a real number $\alpha>0$ and a sequence $(X_n)_{n\ge 1}\subset \mathcal{X}^{T\epsilon}_b(c)$ such that 
\be\label{eq: non conv en proba}
\Pro{|\ell_{T}^{\epsilon}(X_n)|/n\ge 1}\ge \alpha,\quad \forall n\ge 1.
\ee
By definition of $ \Xc_{b}^{T\epsilon}(c)$, there exists   $(\eta_n)_{n\ge 1} \subset  L^{0}(\Fc;\Mb)$ such that $\normeM{\eta_n}\le c$  and $X_n+S_TS_0^{-1} \eta_n \in K^{\epsilon}_T$,  for all $n\ge 1$. Set $\tilde X_n:=X_n/n$ and $\tilde \eta_n:=\eta_n/n$,  so that  $\tilde X_n+ S_TS_0^{-1} \tilde\eta_n \in  K^{\epsilon}_T $ and $ c/n \to 0$. Under \NFLVRe , this implies that  $\ell_{T}^{\epsilon}(\tilde X_n)\to 0$ in probability. This contradicts \reff{eq: non conv en proba}. 
\ep

\begin{Lemma}\label{lem : Lc borne en proba sous RNFLVR} Assume that \RNFLVR~ holds. Then, for all $c\in \R_+$, the set $\{\|L\|_{M(\T\x \bRpd)} \,:\, L \in \Lc_{b}(c)\} \subset L^0(\Fc)$ is bounded in probability.
\end{Lemma}

\proof  Let $\epsilon$ be as in Definition \ref{def: RNFLVR}.

1.    Fix  $L \in \Lc_{b}(c)$    a $c$-admissible strategy   and set 
$$
V^{L\epsilon}_{T}(f):= L\left(G^{\epsilon}_{T}( f)\right)\;,\; f\in \Cb.
$$
Since 
$$
G^{\epsilon}_{T}(f)(s,x,y)=G_{T}(f)(s,x,y)  + \epsilon_s(x,y) (S_{T}(x)/S_{s}(x)) f(x) ,
$$
 it follows that 
\be\label{eq : Veps en fonction V et eps L}
V^{L\epsilon}_{T}(f)=V^{L}_{T}(f)+ \mu^{L}(f),
\ee
where 
 $$
 \mu^{L}(f):=  \int_{\T\x \bRpd} \epsilon_s(x,y) (S_{T}(x)/S_{s}(x)) f(x)dL(s,x,y).
 $$
Since $L$ is $M_{+}(\T\x \bRpd)$-valued,  $\mu^{L}(f) \in L^0(\Fc; \R_+)$  for all $f\in \Cb_+$. This implies that $\Pas$ $\mu^{L} \in \Mb_+ \subset K_{T}$.  Recalling the definition of $\Lc_{b}(c)$, this shows that 
$$
V^{L\epsilon}_{T} +   \frac{S_T}{S_0}\eta  \in K_{T} \;\;\Pas,
$$
for some $\eta \in \Mb$ with $\normeM{\eta} \leq c$. Now observe that $K_{T}\subset K^{\epsilon}_{T}$, and therefore 
$$
V^{L\epsilon}_{T}   \in  \Xc_{b}^{T\epsilon}(c).
$$
In particular, this shows that  
\be\label{eq: Xc inclus dans Xc eps} 
\Xc_{b}^{T}(c) \subset  \Xc_{b}^{T\epsilon}(c).
\ee
 
2.  Let $L\in \Lc_{b}(c)$ be as  above. By  \reff{eq : Veps en fonction V et eps L}  and \reff{eq: def ell par dualite} applied to $\ell_{T}^{\epsilon}$,
 \b*
\ell_{T}^{\epsilon}(V^{L\epsilon}_{T})\ge \ell_{T}^{\epsilon}(V^{L}_{T} )+   \ell_{T}^{\epsilon}(  \mu^{L}).
\e*
Appealing to \reff{eq: Xc inclus dans Xc eps} and Lemma \ref{lem : NFLVR implique borne proba val liqui}, this implies that $\{\ell_{T}^{\epsilon}( \mu^{L}), L\in \Lc_{b}(c)\}$ is bounded in probability. We now apply Remark \ref{rem : ell} to $\ell_{T}^{\epsilon}$: 
  $$
  \ell_{T}^{\epsilon}( \mu^{L})\ge \iota_T \normeM{ \mu^{L}}
  $$ 
 where $\iota_T \in L^{0}(\Fc;(0,\infty))$. Since $L\in M_{+}(\T\x \bRpd)$, the lemma now follows from
 $$
\normeM{\mu^{L}}= L(\epsilon S_T/S) \geq a  \|L\|_{M(\T\x \bRpd)},
 $$
where $a:=\inf \{\epsilon_s(x,y) S_T(x)/S_s(x) \,:\, (s,x,y)\in \T\x \bRpd  \} \in L^0(\Fc; (0,\infty))$ by a continuity and compactness argument, recall Remark \ref{rem : S et lambda adpate et continus} and the definition of $\Upsilon$.
\ep
\\

In order to deduce from the above the required closure property, we now state a version of Koml\`os lemma.  
\begin{Lemma}\label{lem : komlos mesures}  Let $E$ be a compact space and $(\tilde L^{n})_{n\ge 1}\subset L^{0}(\Fc;$ $M_{ +\beta}(E))$ be    bounded in probability.  Then, there exists a  sequence $(\bar L^{n})_{n\ge 1}$, satisfying  $ \bar L^{n}\in {\rm conv}(\tilde L^{k},\;k\ge n)$ for all $n\ge 1$, which   weak*-converges $\Pas$ to some $L\in L^{0}(\Fc; M_{+}(E))$.
\end{Lemma}

\proof a. Let $I:=(f_{k})_{k\ge 1}$ be a dense subset of the separable space $C_\beta(E)$. Then, combining   \cite[Lemma 5.2.7]{Kabanov1}   with a diagonalisation procedure shows that  there exists a  sequence $(\bar L^{n})_{n\ge 1}$ such that $\bar L^{n}\in {\rm conv}(\tilde L^{k},\;k\ge n)$ for all $n\ge 1$, and such that $(\bar L^{n}(f_{k}))_{n\ge 1}$ converges $\Pas$ to some $\zeta_{k} \in L^{0}(\Fc,\R)$.  %
We set 
$L(f_{k})=\zeta_{k}$.
 
b.  We now  extend $L$ to $C(E)$.  To do this, we note that, for each $g\in C(E)$, one can find a sequence $(g_{k})_{k\ge 1}\subset I$ that converges in $C_\beta(E)$ to  $g$. We claim that $\lim_{k\ge 1} L(g_{k})$ is well defined and does not depend on the chosen sequence $(g_{k})_{k\ge 1}$ that converges to $g$.   
First, we show that $(L(g_{k}))_{k\ge 1}$ is   $\Pas$ a Cauchy sequence. Indeed, 
\b*
|L(g_{k})-L(g_{k'})|
&=& 
\lim_{n\to \infty} |\bar L^{n}(g_{k})-\bar L^{n}(g_{k'})| 
\\
&\le& 
\esssup_{n\ge 1} \|\bar L^{n}\|_{M(E)} \|g_{k'}-g_{k}\|_{C(E)} .
\e*
The first term on the right is a.s. bounded while the second term converges to $0$ as $k,k'\to \infty$, since $ C_\beta(E)$ is complete. It remains to check that the result is the same if we consider two different approximating sequences. But this follows immediately from the same estimates. For $g$ as above, we can then define $L(g):=\lim_{k\ge 1} L(g_{k})$.

c. To see that  $(\bar L^{n})_{n\ge 1}$ converges $\Pas$ to $L$ in the weak* topology, let us note that, for $g\in C(E)$, one has 
\b*
|\bar L^{n}(g)-L(g)|
&\le& 
|\bar L^{n}(g_{k})-L(g_{k})| +  2 \sup_{m\ge 1} \|\bar L^{m}\|_{M(E )}    \|g-g_{k}\|_{C(E)} 
\e*
Taking $(g_{k})_{k\ge 1}$ that converges to $g$ in $ C_\beta(E)$ leads to the required result by first taking the limit $n \rightarrow \infty$, and then $k \rightarrow  \infty$.

d. The above also shows that the map $C_\beta(E) \ni g \mapsto L(g)$ is continuous $\Pas$ The linearity is obvious.  

e. The measurability is obvious since $L(f_{k})$ is $\Fc$-measurable as the $\Pas$ limit of $\Fc$-measurable random variables, which extends to $L(g)$ for any $g$ by the construction in b.~above.   
\ep

\begin{Corollary}\label{cor : komlos mesure de transfer} Let $(  L^{n})_{n\ge 1} \subset \Lc$ be such that $(  L^{n}_{T})_{n\ge 1}$ is bounded    in probability. Then, there exists a  sequence $(\bar L^{n})_{n\ge 1}$, satisfying $ \bar L^{n}\in {\rm conv}( L^{k},\;k\ge n)$ for all $n\ge 1$,  that converges $\Pas$ for the weak* topology to some $L\in  \Lc$. 
\end{Corollary}
 
\proof It suffices to apply  Lemma \ref{lem : komlos mesures} to $E:=\T\x \bRpd$. The weak*-measurability property of Definition \ref{def: trans measure} follows by the weak*-convergence property of Lemma \ref{lem : komlos mesures}.  \ep\\
 
 We are now in position to conclude the proof of Theorem \ref{thm: fermeture Xb} by   using routine arguments, which we provide here for completeness. \\
{\bf Proof of Theorem \ref{thm: fermeture Xb}.} a. Let us suppose that $(X_n)_{n\ge 1}\subset \Xc_b^T$  weak*-converges $\Pas$ to $X\in L^{0}(\Fc_T; \Mb)$. Moreover, assume that  there exists $\eta_n\in L^{0}(\Fc_T; \Mb)$ such that $X_n+ S_T S_0^{-1} \eta_{n}\in K_T$ a.s. and $c:=\sup_n\normeM{\eta_n} \in L^{\infty}$. Let $(L^{n})_{n\ge 1} \in \Lc_{b}(c)$ be a sequence of transfer measures associated to $(X_{n})_{n\ge 1}$, i.e. such that 
\be\label{eq: mun le LnGf}
X_{n}(f)\le L^{n}(G_{T}(f))\;\text{ for all $n\ge 1$ and $f \in \Cb_{+}$}.
\ee
  It follows from Lemma \ref{lem : Lc borne en proba sous RNFLVR}  that $(L^n)_{n\ge 1}$ is bounded in probability. Applying Corollary \ref{cor : komlos mesure de transfer}, we may assume without loss of generality (up to passing to  convex combinations) that  $   L^n_T $   weak*-converges  $\Pas$ to some $  L\in \Lc_{b}$.    Using Remark \ref{rem : S et lambda adpate et continus}, one easily checks that    $L^{n}(G_{t} {(f)})\to L (G_{t}(f))$ $\Pas$ for all $f \in \Cb$. Passing to the limit in \reff{eq: mun le LnGf} thus implies  $X(f)\le L(G_{T}(f))$ for all $f \in \Cb_+$.  This shows that $\mathcal{X}_b^T$ is  Fatou-closed. 

b. {By Krein-\v{S}mulian's Theorem, (c.f. Corollary, Ch. IV, Sect. $6.4$ of \cite{Schaefer})}, it suffices to show that  $ \hat \Xc_{b}^{T}\cap B_{1}$ is $\sigma( L^{\infty}(\Fc_T; \Mb), L^{1}(\Fc_T;\Cb))$-closed, where $B_{1}$ is the unit ball of  $L^{\infty}(\Fc_T; \Mb)$. To see this, let $(\hat X_\alpha)_{\alpha \in \Ic}$ be a net in  $ \hat \Xc_{b}^{T}\cap B_{1}$  which converges  $\sigma( L^{\infty}(\Fc_T; \Mb), L^{1}(\Fc_T;\Cb))$ to some $\hat  X\in B_1$. After possibly passing to convex combinations, we can then construct a sequence $(\hat X_n)_{n\ge 1}$ in $ \hat \Xc_{b}^{T}\cap B_{1}$ which weak*-convergences $\Pas$ to  $\hat X$, see e.g. \cite[Lemma 4.1]{BTa2010}. By the continuity property of Remark \ref{rem : S et lambda adpate et continus}, this implies that 
$(  X_n)_{n\ge 1}$ in $   \Xc_{b}^{T} $  weak*-converges $\Pas$ to  $  X$, with $X_{n}(f):=\hat X_{n}(f  S_{T}/S_{0} )$ and $X(f):=\hat X (f  S_{T}/S_{0} )$. Since $(\hat  X_{n})_{n\ge 1}\subset B_1$, one easily checks that $( X_{n})_{n\ge 1}$ is indeed Fatou-convergent. Since $\Xc_{b}^{T}$ is   Fatou-closed, this shows that $ \hat X\in \hat \Xc_{b}^{T}$.
 \ep

\section{Equivalence with the existence of a strictly  consistent price system}

 From now on,  we define the set of {\sl strictly consistent price systems},  $\Mc({\rm int}(\tdual{K}))$,
as    the set of $\Cb$-valued weakly $\F$-adapted c\`adl\`ag processes $Z=(Z_{t})_{t\in \T}$ such that 
\begin{enumerate}[({\rm \bf Z}a.)]
\item\label{ite : Ztau dans Ktau} $Z_{\tau}\in {\rm int}(\tdual{K}_{\tau})$ $\Pas$ for all $\tau \in \Tc$,
\item\label{ite : Ztau- dans Ktau} $Z_{\tau-}\in {\rm int}(\tdual{K}_{\tau})$ $\Pas$ for all predictable $\tau \in \Tc$,
\item\label{ite : ZS mart Cb valued}    $Z S/S_0$ is a $\Cb$-valued  martingale satisfying $\normeC{ZS/S_{0}}\in L^{1}$.
 \end{enumerate}
 
The terminology {\sl strictly consistent price systems} was introduced in \cite{schach04}. They play the same role as equivalent martingale measures in frictionless markets, see e.g.\cite{Kabanov1}.

 \begin{Remark} {\rm 
  Proposition \ref{prop: int  K'} and  Proposition \ref{prop: propriete K et K'} allows to give a sense to the assertions ({\rm \bf Z}\ref{ite : Ztau dans Ktau})  {and} ({\rm \bf Z}\ref{ite : Ztau- dans Ktau}). 
}
\end{Remark}

\subsection{Existence under (RNFLVR)}

The main result of this section extends the first implication in \cite[Theorem 1.1]{GLR} to our setting.

 \begin{Theorem}\label{thm: existence Z sous RNFLVR} Let \RNFLVR~hold. Then, there exists $\epsilon \in \Upsilon$ such that $\Mc({\rm int}(\tdual{K}))\supset\Mc({\rm int}(\Kdeps))\ne \emptyset$.
\end{Theorem}

In order to show the above, we shall follow the usual Hahn-Banach separation argument based on the weak*-closure property of Theorem  \ref{thm: fermeture Xb} above. This is standard but requires special care in our infinite dimensional setting. In particular, we shall first need to show that simple strategies are admissible.   To this purpose, we introduce the notation 
\be\label{eq: hat K tau}
\hat K_{\tau}:=\{  S_0 /S_{\tau} \nu : \nu \in K_{\tau} \} \;\mbox{ for }\tau \in \Tc. 
 \ee
Clearly, the measurability of Proposition \ref{prop: propriete K et K'} extends to $\hat K$.  An element of $-\hat K_{\tau}$ can be interpreted as a portfolio holding, evaluated in terms of time-$0$  prices, obtained by only performing immediate transfers at time $\tau$. The following technical result is obvious in discrete time settings.  

\begin{Proposition}\label{prop : PropFondSurX V2}  $L^{\infty}(\mathcal{F}_{\tau};-\hat K_{\tau}) \subset \hat \Xc_{b}^{T}$ for all $\tau \in \Tc$. 
\end{Proposition}
\proof  
Fix $\hat \xi \in L^{\infty}(\mathcal{F}_{\tau};-\hat	 K_{\tau}).$
We must show that there exists   $L\in \Lc_{b}$ such that 
\be\label{eq: representation par L}
V_{T}^{L}=\frac{S_T}{S_\tau}\xi \;\;\;\mbox{ where } \xi := (S_\tau/S_0)\hat \xi.
\ee
This equation is satisfied if  the portfolio process $V^{L}$ satisfies
$$
V_{\tau}^{L}(g)= L_{\tau}(H(1 \otimes g))= \xi(g), \text{ for all } g\in \Cb,
$$
$V_{t}^{L}=0 \text{ on } \{t < \tau\}  \text{ and } V_{t}^{L}=\frac{S_t}{S_\tau}\xi  \text{ on } \{t \geq \tau\}$. Equivalently the random measure $\mu:=- L \circ H$  shall satisfy $\mu(f)=-\xi(f(\tau,\cdot))$ for $f \in C([0,T]\x \bRp)$, i.e.
\be\label{eq: representation par L 2}
 \mu=-\delta_\tau \otimes \xi.
\ee
We can now apply Corollary \ref{cor: measurable selection L} in the Appendix and define $L$ by
\begin{equation} \label{eq: representation par L 3}
L(\omega)=J(\lambda(\omega),\mu(\omega)).
\end{equation}
Since  $\lambda \1_{[0,t] \x \bRpd}$ and $\mu\1_{[0,t] \x \bRp}$ are $\Fc_t$-measurable, it follows that $L$ has the properties required by Definition \ref{def: trans measure}, recall Remark \ref{rem : S et lambda adpate et continus} and (a.) of Proposition \ref{prop : joint measurablity} in the Appendix. As $\hat \xi \in L^\infty(\Fc; \Mb_\beta)$, the strategy is bounded in the sense of Definition \ref{def: t-lower bound}
\ep\\

We can now provide the proof of  {Theorem \ref{thm: existence Z sous RNFLVR}.}

\noindent{\bf Proof of Theorem \ref{thm: existence Z sous RNFLVR}.} Fix $  \epsilon\in \Upsilon$ such that {\NFLVR}$^{  \epsilon}$~holds.  We shall construct $Z$ such that ({\rm \bf Z}\ref{ite : ZS mart Cb valued}) holds  and $Z_{\tau}\in   \Kdeps_{\tau}$ for all stopping times $\tau \in \Tc$.  In particular, as a martingale, $Z S/S_{0}$ has to be  c\`adl\`ag (cf. \cite[Ch. II, Th. (2.9)]{Revuz-Yor}), and, since $S$ has  continuous paths and takes strictly positive values,  $Z$ is c\`adl\`ag.   We shall also show that $Z_{T}\ge 0$ and that $Z_{T}(\bar x)>0$ for at least one $\bar x\in \bRp$ (actually along a dense sequence). Since  $(Z S/S_0)(\bar x)$ is a  martingale, this implies that $Z_{\tau}(\bar x)>0$ for all stopping times $\tau \in \Tc$. In view of the definition of $\Kdeps_{\tau}$ this readily implies that   $Z_{\tau}\in {\rm int}(\tdual{K}_{\tau})$. Our continuity assumptions, see Remark \ref{rem : S et lambda adpate et continus}, then imply that   $Z_{\tau-}\in   \Kdeps_{\tau}$ for all predictable stopping time $\tau \in \Tc$. Similarly as above, we must have $Z_{\tau-}(\bar x)>0$, see e.g. \cite[Lemma 2.27]{JS},  so that $Z_{\tau-}\in {\rm int}(\tdual{K}_{\tau})$, whenever $\tau$ is  predictable. This will show that $\Mc({\rm int}(\tdual{K}))\ne \emptyset$. To find an $\bar \epsilon \in \Upsilon$ such that $\Mc({\rm int}(\Kdepsb))\ne \emptyset$, we just note that \RNFLVR~for the original transaction costs $\lambda$ implies   \RNFLVR~for some $\lambda^{\bar \epsilon}$ defined as in \reff{eq: lambda epsilon} for some $\Upsilon \ni \bar \epsilon <  \epsilon$. This $\bar \epsilon$ can be easily constructed by using the argument of Remark \ref{rmk: ex lambda epsilon}.

1. It follows from the assumption \NFLVRe~ that $\hat \Xc_{b}^{T\epsilon} \cap L^{\infty}(\Fc_T; \Mb_+)=\{0\}.$
 The Hahn--Banach  theorem and Theorem \ref{thm: fermeture Xb} then imply that,  for any $\nu \in L^{\infty}(\Fc_T; \Mb_{+})\setminus\{0\}$, there exists $f_{\nu}\in L^1(\Fc_T; \Cb)$ and a real constant $a_{\nu}$ such that 
\be \label{eq: Separ} 
\Esp{X(f_{\nu})}<a_{\nu}<\Esp{\nu(f_{\nu})},\quad \forall X\in  \hat \Xc_{b}^{T\epsilon} \cap L^{\infty}(\Fc_T; \Mb).
\ee
Since $\hat \Xc_{b}^{T\epsilon}$ is a cone of vertex $0$ which contains $L^{0}(\Fc_{T};-\Mb_{+})$, we deduce that 
\be
&f_{\nu}\in L^1(\Fc_T; \Cb_{+})\label{eq: fnum ge 0}&\\
\label{eq: EX(fnu)le 0}
&a_{\nu} > 0 \;\mbox{ and }\; \Esp{X(f_{\nu})}\le 0 \;\mbox{ for all } X\in \hat \Xc_{b}^{T\epsilon} \cap L^{\infty}(\Fc_T; \Mb).&
\ee   
Also observe that we may assume without loss of generality that $ \normeC{f_{\nu}}\le 1$.

2. In the following, we use the fact that  $\Mb_+$ is the $\sigma(\Mb,\Cb)$-closure  of the cone generated by the countable basis $(\delta_{x_{k}})_{k\ge 1}$, where $(x_{k})_{k\ge 1}=\Q_{+}\cup \{\infty\}$.  We set $A_k(\nu):=\{\omega \in \Omega : \delta_{x_{k}}(f_{\nu})(\omega)>0\}$ for $\nu \in  L^{\infty}(\Fc_T; \Mb_{+})  \backslash \{0\}$ and
$$
\Ac_{k}:=\left\{ A_k(\nu) : \nu \in  L^{\infty}(\Fc_T; \Mb_{+})  \backslash \{0\}\right\},\quad k\in \N.
$$
 If $\Gamma \in \mathcal{F}_{T}$ is a non-null set, then
$\Pro{\Gamma \cap  A_k(\nu)}>0$ for $\nu$   defined by $\nu:=\delta_{x_{k}}\1_{\Gamma}\in
L^{\infty}(\Fc_t; \Mb_+)$. This follows from the left-hand side of \reff{eq: EX(fnu)le 0} and the right-hand side of  \reff{eq: Separ}. By virtue of  \cite[Lemma 2.1.3 p74]{Kabanov1}, we
can then, for $k$ given, find  a countable subfamily $\{A_k(\nu_{k}^{i}): i\in \N \} \subset \Ac_{k}$ such that 
\be\label{eq: Bk full measure + partition}
B_{k}:= \bigcup_{i\in \N}A_k(\nu_{k}^{i})  \mbox{ satisfies } \Pro{B_{k}}=1.
\ee
Therefore, $B:=\cap_k B_{k}$ is a  set of measure $1$.

Let us set  
$$
\check Z_T:=\sum_{k,i\ge 1} 2^{-k-i}f_{\nu_{k}^{i}}.
$$ 
On each $B_{k}$,
$\check Z_T(x_{k})>0$. This follows from \reff{eq: Bk full measure + partition} and \reff{eq: fnum ge 0}. Since $x\mapsto \check Z_{T}(x)$ is continuous, this implies that $\check Z_{T}(x)\ge 0$ for all $x\in \bRp$ $\Pas$
For later use, note that 
\be\label{eq: EX(Z)le 0}
  \Esp{X(\check Z_{T})}\le 0 \;\mbox{ for all } X\in \hat \Xc_{b}^{T\epsilon} \cap    L^{\infty}(\Fc_T; \Mb),
\ee 
by \reff{eq: EX(fnu)le 0} and the definition of $\check Z_{T}$.

3. Let  $\Mb^{1}$ be the closed unit ball of $\Mb$, i.e. $\Mb^{1}:=\{\eta \in \Mb: \normeM{\eta} \le 1\}$. Given $\tau \in \Tc$, we set $Z_{\tau}:=\Esp{ \check Z_{T}|\Fc_{\tau}}S_0/S_{\tau}$.  We now show that $Z_{\tau} \in  \Kdeps_{\tau}$.   
Indeed, if it is not the case then, for every $\omega$ in the non-null set $\Lambda_{\tau}:=\{ Z_{\tau}\notin   \Kdeps_{\tau} \}\in \mathcal{F}_{\tau}$, we may find $\xi_{\omega}\in K_{\tau}^{\epsilon}(\omega)\cap \Mb^{1}$ such that $\xi_{\omega}(Z_{\tau})<0$.   It follows that the set 
$$
\Gamma:=\left \{(\omega,\xi)\in \Omega\times \Mb^{1}:~\xi \in K_{\tau}^{\epsilon}(\omega)\mbox{ and }  \xi(Z_{\tau}(\omega))<0 \right \}
$$
is of full measure on $\Lambda_{\tau}\times \Mb^{1}$, i.e.  $\Lambda_{\tau}\setminus \{\omega \in \Omega: \exists \; \xi \in \Mb^{1}$ s.t. $(\omega,\xi)\in \Gamma\}$ $=$ $\emptyset$ up to $\P$-null sets. As $\Gamma$ is $\Fc_{\tau}\otimes \Bc(\Mb_{\sigma})$-measurable,  by a measurable selection argument, we then obtain an $ \mathcal{F}_{\tau}$-measurable selector $\xi$ such that $(\omega,\xi(\omega))\in \Gamma$ on $\Lambda_{\tau}$ and $\xi=0$ otherwise, see e.g.~\cite[Theorem 5.4.1]{Kabanov1} or \cite[Theorem 18.26]{Aliprantis-Border}. One has $\Esp{-\xi(Z_{\tau})}>0$. Suppose for the moment that 
\be\label{eq : claim E xi Z tau}
\Esp{-\xi(Z_{\tau})}= \Esp{-\xi(   \check Z_{T}S_0/S_{\tau})} .   
\ee
Then, since  $\{  (S_0 /S_{\tau}) \nu : \nu \in L^{\infty}(\mathcal{F}_{\tau}; -K_{\tau}^{\epsilon})\}  \subset \hat \Xc_{b}^{T\epsilon}$,    see Proposition \ref{prop : PropFondSurX V2}  above, we obtain a contradiction to \reff{eq: EX(Z)le 0} if $\tau$ is such that  $ \normeC{S_0/S_{  \tau}}\in L^{\infty}$.  This shows that $Z_{\tau} \in  \Kdeps_{\tau}$ for such stopping times $\tau$. In view of \reff{eq : Hyp  prix posi conti} and \reff{eq : Hyp prix conti en temps}, the general case is obtained by a standard localization argument. 

4. It remains to prove \reff{eq : claim E xi Z tau}. We notice that the $\xi$ in \reff{eq : claim E xi Z tau} is  $\Fc_{\tau}$-measurable, by construction.  Thus, the random measure 
 $ (S_0/S_{\tau})\xi$  can   be viewed as an optional random measure with respect to $(\Fc_{t\vee \tau})_{t\in \T}$. Since  $Z_{\tau}S_{\tau}/S_\tau$  is by construction the $(\Fc_{t\vee \tau})_{t\in \T}$-optional projection at the stopping time $\tau$ of  $Z_{T}S_T/S_\tau$ $=$ $\check Z_{T}S_{0}/S_{\tau}$, it follows from Theorem \ref{thm : ProjectOpt} in the Appendix that 
$$
\Esp{\xi(Z_{\tau})}=\Esp{\xi(Z_{\tau}S_{\tau}/S_{\tau})}= \Esp{\xi(Z_{T}S_{T}/S_{\tau})}=   \Esp{\xi(\check Z_{T}S_0/S_{\tau})}. 
$$
     \ep

 \subsection{Existence of strictly consistent price systems implies (RNFLVR)}
 
 The fact that the existence of strictly consistent price systems implies (NFLVR) follows as usual from the  super-martingale property of admissible wealth processes  when evaluated along   consistent price systems.

In our infinite dimensional setting, this super-martingale property can not be deduced directly from an integration by parts argument as in e.g.~\cite{C-S}. We instead appeal to an optional projection theorem which we state in the Appendix.
 \\
 
 In the following, we let $\Mc( \tdual{K})$ be defined  as $\Mc( {\rm int} (\tdual{K}))$ at the beginning of Section \ref{sec : NFLVR} but with $\tdual{K}$ in place of ${\rm int}({\tdual{K}})$.
 
\begin{Proposition}\label{prop: VZ super mart} Fix $Z\in \Mc(  \tdual K )$ and $L\in \Lc_{b}$. Then, $(V^{L}_{t}(Z_{t}))_{t\in \T}$ is a super-martingale. 
\end{Proposition}

\proof Fix $t\ge s  \in \T$ and $L\in   \Lc_{b}$.

1. Fix $\tau \in \Tc$ and assume that      
$$
\mu^{L}_{S,\tau}(f)=\int_{[0,\tau)\x \bRpd} \left[(fS_0/S_{u})(y)+(fS_0/S_{u})(x)\right](1+\lambda_{u}(x,y))dL(u,x,y)\;,\; f\in \Cb,
$$ 
satisfies
\be\label{eq : borne hat L}
  \normeM{\mu^{L}_{S,\tau}}\in L^{\infty}. 
\ee
In the following, we write $X^{\tau}$ for the stopped process $X_{\cdot \wedge \tau}$ associated to an adapted process $X$ taking values in  $\Cb$, $C(\bRpd)$ or $\Mb$.  One has 
\b*
V^{L,\tau}_{t}(Z_{{t}}^{\tau})= A^{\tau}_{t }-B^{\tau}_{t} + \Delta_{t}^{\tau}
\e*
where 
\b*
A^{\tau}_{t}&:=& \int_{[0,t\wedge \tau)\x \bRpd}  (Z^{\tau}_{t} S^{\tau}_{t}/S_{u})(y)dL(u,x,y)\\
\\
B^{\tau}_{t}&:=&\int_{[0,t\wedge \tau)\x \bRpd} (Z^{\tau}_{t}S^{\tau}_{t}/S_{u})(x)(1+\lambda_{u}(x,y)) dL(u,x,y)\\
\Delta_{t}^{\tau}&:=& \int_{\{t\wedge \tau\}\x \bRpd}\left\{ (Z^{\tau}_{t} S^{\tau}_{t}/S_{u})(y)-(Z^{\tau}_{t}S^{\tau}_{t}/S_{u})(x)(1+\lambda_{u}(x,y))\right\} dL(u,x,y).
\e*
First note that  
$$
\Delta_{t}^{\tau}=\int_{\{t\wedge \tau\}\x \bRpd}\left\{  Z_{t\wedge \tau} (y)- Z_{t\wedge \tau} (x)(1+\lambda_{t\wedge \tau}(x,y))\right\} dL(u,x,y) \le 0 
$$
since $Z_{t\wedge \tau}\in L^{0}(\Fc_{t\wedge \tau};\tdual{K}_{t\wedge \tau})$, recall \reff{eq: dual cone in  C(K)}. Hence
\be\label{eq: V le A-B}
V^{L,\tau}_{t}(Z_{{t}}^{\tau})\le A^{\tau}_{t }-B^{\tau}_{t}.
\ee
Since $\normeC{ZS/S_{0}}\in L^{1}$, \reff{eq : borne hat L}  imply that $A^{\tau}_{t},B^{\tau}_{t}\in L^{1}$. Moreover, $Z$, $S$ and $\lambda$ take non negative values and the  $(\Fc_{s\vee u})_{u\in \T}$-optional projections of $Z^{\tau}_{t} S^{\tau}_{t}/S_{\cdot}$ and  $(Z^{\tau}_{t} S^{\tau}_{t}/S_{\cdot})  (1+\lambda_{\cdot}(x,y))$ are  $(Z^{\tau}S^{\tau})_{(s\vee \cdot)\wedge t}/S_{\cdot}$ and $((Z^{\tau}S^{\tau})_{(s\vee \cdot)\wedge t}/S_{\cdot})$ $(1+\lambda_{\cdot}(x,y))$ since $ZS$ is a martingale, and the other processes are adapted and continuous (and therefore optional).  Applying Theorem \ref{thm : ProjectOpt}   in the Appendix, we get that $\Esp{A^{\tau}_{t}|\Fc_{s}}-\Esp{B^{\tau}_{t}|\Fc_{s}} =\alpha^{\tau}_s-\beta^{\tau}_s$ where
\b*
\alpha^{\tau}_s&:=& \int_{[0,s\wedge \tau]\x \bRpd}  ( Z_{s} S_{s}/S_{u})(y)dL(u,x,y)\\
&&+\Esp{\int_{(s,t\wedge \tau)\x \bRpd}   Z_{u}(y) dL(u,x,y)|\Fc_{s}}\\
\\
\beta^{\tau}_s&:=&\int_{[0,s\wedge \tau]\x \bRpd}  ( Z_{s} S_{s}/S_{u})(x)(1+\lambda_{u}(x,y)) dL(u,x,y)\\
&&+\Esp{\int_{(s,t\wedge \tau)\x \bRpd}   Z_{u}(x) (1+\lambda_{u}(x,y))  dL(u,x,y)|\Fc_{s}}. 
\e*
Moreover, 
$$
 \int_{[0,s\wedge \tau]\x \bRpd}\left[  ( Z_{s} S_{s}/S_{u})(y)-( Z_{s} S_{s}/S_{u})(x)(1+\lambda_{u}(x,y)) \right] dL(u,x,y)=V^{L,\tau}_{s}(Z^{\tau}_{s}),
$$ 
while $ Z_{u}(y) - Z_{u}(x) (1+\lambda_{u}(x,y))\le 0$, since $Z_{u}\in \tdual{K}_{u}$. Since $L$ is a positive random measure, the above combined with \reff{eq: V le A-B} implies that  
$$
\Esp{V^{L,\tau}_{t}(Z_{t})|\Fc_{s}}\le \Esp{A^{\tau}_{t}-B^{\tau}_{t}|\Fc_{s}}\le V^{L,\tau}_{s}(Z^{\tau}_{s}). 
$$

2. We now turn to the general case. In view of Remark \ref{rem : S et lambda adpate et continus}, we can find an increasing sequence of stopping times $(\tau_{n})_{n\ge 1}$ such that $\tau_{n}\to \infty$ $\Pas$ and 
$$
\sup_{u \in \T} ( \normeC{S_0/S_{u\wedge \tau_{n}}}+\|\lambda_{u\wedge \tau_{n}}\|_{C(\bRpd)}+\|L^{-}_{u\wedge \tau_{n} }\|_{M(\bRpd)}) \in L^{\infty},
$$
in which $L^{-}_{u}= \1_{[0,u)\x \bRpd} L$. 
Then, \reff{eq : borne hat L} holds for each $n\ge 1$, and therefore 
$$
\Esp{V^{L,\tau_{n}}_{t}(Z^{\tau_{n}}_{t})|\Fc_{s}} \le V^{L,\tau_{n}}_{s}(Z^{\tau_{n}}_{s})
$$
or equivalently  
\be \label{IneSuper}
\Esp{V^{L,\tau_{n}}_{t}(Z_{t}^{\tau_{n}})^{+}|\Fc_{s}} \le \Esp{V^{L,\tau_{n}}_{t}(Z_{t}^{\tau_{n}})^{-}|\Fc_{s}}+ V^{L,\tau_{n}}_{s}(Z_{s}^{\tau_{n}}),
\ee
in which the superscripts $^{+}$ and $^{-}$ denote the positive and the negative parts. 
Moreover, the definition of $\Lc_{b}$ implies that 
there exists $\eta \in \Mb$ such that $\normeM{\eta}\le c$, for some $c\in \R_+$, for which   
$$
V^{L,\tau_{n}}_{t}+(S^{\tau_{n}}_{t}/S_{0})\eta\in K_{t\wedge\tau_{n}}.
$$
Since $Z^{\tau_{n}}_{t}\in \tdual{K}_{\tau_{n}\wedge t}$, it follows that 
$$
V^{L,\tau_{n}}_{t}(Z_t^{\tau_{n}})+\eta(S_{t}^{\tau_{n}}Z_t^{\tau_{n}}/S_{0})\ge 0.
$$
 Therefore, $V^{L,\tau_{n}}_{t}(Z_t^{\tau_{n}})^{-}\le |\eta(S_{t}^{\tau_{n}}Z_t^{\tau_{n}}/S_{0} )|$. On the other hand,    $\eta(S_{t}^{\tau_{n}}Z_t^{\tau_{n}}/S_{0} )=\eta \left(\Esp{S_TZ_T/S_0|\Fc_{\tau_{n}\wedge t}}\right)$ $=$ 
$\Esp{\eta \left(S_TZ_T/S_0\right)|\Fc_{\tau_{n}\wedge t}}$ by Proposition \ref{ProjectOpt-Equal} in the Appendix and ({\rm \bf Z}\ref{ite : ZS mart Cb valued}), which implies that the sequence $(\eta(S_{t}^{\tau_{n}}Z_t^{\tau_{n}}/S_{0}) )_{n\ge 1}$ is uniformly integrable and so does $(V^{L,\tau_{n}}_{t}(Z_t^{\tau_{n}})^{-})_{n\ge 1}$.  Since the later converges a.s.~to $V^{L}_{t}(Z_t)^{-}$ as $n\to \infty$,  it follows that  $\Esp{V^{L,\tau_{n}}_{t}(Z_{t}^{\tau_{n}})^{-}|\Fc_{s}}$ converges a.s.~to $\Esp{V^{L}_{t}(Z_{t})^{-}|\Fc_{s}}$.     It is then sufficient to apply   Fatou's Lemma to the left-hand side of   (\ref{IneSuper}) to deduce that
\b*
\Esp{V^{L}_{t}(Z_{t})^{+}|\Fc_{s}} \le \Esp{V^{L}_{t}(Z_{t})^{-}|\Fc_{s}}+ V^{L}_{s}(Z_{s}), 
\e*
which concludes the proof. 
 \ep

 \begin{Corollary}\label{cor : M ne empty imply NFLVR} Assume there exists  $Z\in \Mc( \tdual{K}) $ such that $Z_{T}(0)>0$. Then, there exists $\Q\sim \P$ such that  $\E^{\Q}[\ell_{T}(X)]\le 0$ for all $X\in \Xc_{b}^{T}$.  In particular, \NFLVR~holds.  
 \end{Corollary}
 
 \proof 1. Let $L\in \Lc_{b}$ be such that $V^{L}_{T}-X\in  K_{T}$ $\Pas$  Then, 
 $V^{L}_{T}(Z_{T})\ge X(Z_{T})$ since $Z_{T}\in \tdual{K}_{T}$. Since $L\in \Lc_{b}$, Proposition \ref{prop: VZ super mart} implies that 
 $$
 \Esp{V^{L}_{T}(Z_{T})}\le V^{L}_{0}(Z_{0})=\int_{\{0\}\x\bRpd} \left(Z_{0}(y)-Z_{0}(x)(1+\lambda_{s}(x,y))\right)dL(s,x,y)\le 0
 $$
 where the last inequality follows from the fact that $Z_{0}\in \tdual{K}_{0}$. We now use \reff{eq: def ell par dualite} and the fact that  $Z_{T}(0)>0$ $\Pas$ to obtain 
 \be\label{eq : XZ ge Z(0)ellX}
 Z_{T}(0) \ell_{T}(X) \le Z_{T}(0) X(Z_{T}/ Z_{T}(0)) = X(Z_{T}), 
 \ee
 so that, by the above, 
\be\label{eq : EXZ le 0}
 \alpha_{Z}\; \E^{\Q}[\ell_{T}(X)]\le  \Esp{X(Z_{T})}  \le 0 
 \ee
 where 
 $$
 d\Q/d\P:=Z_{T}(0)/\alpha_{Z}\;\mbox{ with }\;\alpha_{Z}:=\Esp{Z_{T}(0)}>0.
 $$
 
 2.  Let  $(X_n,c_{n})_{n\ge 1}  \subset \Xc_{b}^T\x  \R_+$ be such that  $\lim_n  c_n=0$ and $X_{n}\in  \Xc_{b}^T(c_{n})$ for all $n\ge 1$. 
 Let $(\eta_{n})_{n\ge 1}\subset \Mb$ be such that $\normeM{\eta_{n}}  \leq c_{n} $ and $X_{n}+\eta_{n}((S_{T}/S_{0})\cdot ) \in K_{T}$ for all $n\ge 1$. 
 Then, 
 $$
  X_{n}(Z_{T})+ \eta_{n}(Z_{T}S_{T}/S_{0} )\ge 0.
 $$
 Since $\eta_{n}(Z_{T}S_{T}/S_{0} )\to 0$ $\Pas$, the last inequality combined with \reff{eq : EXZ le 0} applied to $X=X_{n}$ implies that $X_{n}(Z_{T})\to 0$ $\Pas$ We conclude from \reff{eq : XZ ge Z(0)ellX} and the fact that $Z_{T}(0)>0$ $\Pas$ that  $\limsup_{n} \ell_{T}(X_{n})\le 0$.
 \ep
 \\
 
 The reciprocal of Theorem 3.2 follows.
\begin{Theorem}  \label{thm: intM non empty implies RNFLVR} Assume that $\Mc(\interior(\Kdeps))\ne \emptyset$ for some $\epsilon \in \Upsilon$,   then \RNFLVR~holds.
\end{Theorem}

 \proof Fix $Z \in \Mc(\interior(\Kdeps))$. In particular, $Z \in \Mc(  \Kdeps )$  and $Z_{T}(0)>0$.  Applying Corollary \ref{cor : M ne empty imply NFLVR} to $\lambda^{\epsilon}$ in place of $\lambda$ implies that \NFLVRe~holds.    
 \ep
 
  \begin{Remark}   {\rm (i).}   The existence of  $Z \in \Mc(\interior(\tdual{K}))$ also implies a  version of the robust no free lunch condition which is weaker than the one of Definition \ref{def: RNFLVR}. More precisely, it  implies that we can find $\epsilon$, satisfying all the conditions in the definition of $\Upsilon$ except that the process $t\mapsto \epsilon_{t}$ may no more be strongly continuous but only c\`adl\`ag, such that \NFLVRe~holds. 
 It is given by 
 $$
   \epsilon_{t}(x,y):=(1+\lambda_{t}(x,y))-  Z_{t}(y)/Z_{t}(x).
 $$ 
 Then, $Z\in \Mc( \Kdeps)$ and $Z_{T}(0)>0$  by construction. 
 To check that the property \NFLVRe~holds, it then suffices to observe that the strong continuity assumption on the process $\lambda$ is not used in the proof of Corollary \ref{cor : M ne empty imply NFLVR}. 
 
 {\rm (ii).} Combining Theorems \ref{thm: existence Z sous RNFLVR} and \ref{thm: intM non empty implies RNFLVR} leads to:  $\Mc(\interior(\Kdeps))\ne \emptyset$ for some $\epsilon \in \Upsilon$ $\Leftrightarrow$  \RNFLVR~holds. One may want to prove:  $\Mc(\interior(\tdual{K}))\ne \emptyset$  $\Leftrightarrow$  \RNFLVR~holds. Actually, Theorem   \ref{thm: existence Z sous RNFLVR} provides the direction $\Leftarrow$. To prove the reverse implication, one will typically need to construct some $\epsilon$ as in (i) above. But this one does not, in general, belong to $\Upsilon$ if one only knows that $Z$ is $\interior(\tdual{K})$-valued. One would need more information,  for instance that  $Z$ is strongly continuous. As a matter of  {fact}, the last equivalence can, in general, only hold if one can remove the strong time continuity condition in the definition of $\Upsilon$, i.e.~deal  with jumps in the bid-ask prices.   As explained in the introduction, we leave this case for further research. 
  \end{Remark}

\section{Appendix}

We report here on technical results that were used in the previous proofs.

\subsection{On optional projections and the measurability of composition of maps}

 We first  provide two standard results, which we  adapt to our context. The  proofs follow  classical arguments and are reported only for completeness.   
 
\begin{Theorem} \label{thm : ProjectOpt} Let $ \T\x \bRp\x  \Omega \ni (t,x,\omega)   \mapsto X_{t}(x)(\omega) \in \R$ be a $\Bc([0,T]\x \bRp)\otimes \Fc/\Bc(\R)$-measurable function,  such that $|X_{t}(x)|\in L^{1}$ for all $(t,x)\in \T\x\bRp$. Let  $\mu\in L^0(\Fc;M(\T\x \bRp))$ be such that $(\mu( [0,t]\x A))_{t\in \T}$ is optional  for any $A\in \Bc(\bRp)$.  Assume further that $|\mu|(|X|)\in L^{1}$.  Then 
\be \label{ProjectOpt-Equal} 
\Esp{\mu(X)}=\Esp{\mu(X^o)},
\ee
where $X^{o}$ is defined as the point-wise optional projection of $X$: 
$$
X^{o}_{t}(x):=\Esp{X_{t}(x)|\Fc_{t}} \;\mbox{ for } x \in \bRp \mbox{ and } t\in \T.
$$ 
\end{Theorem}
\proof Obviously, one can restrict to the case where $\mu$ is non-negative by considering separately $\mu^{+}$ and $\mu^{-}$. If $X$ is of the form $X_{t}(x)(\omega)=1_{A}(x)\xi_{t}(\omega)$ with $A\in \Bc(\bRp)$ and $\xi$ is $\Fc\otimes \Bc([0,T])$-measurable and bounded, then the optional projection $X^{o}$ of $X$ is given by  $1_{A}(x)\xi^{o}_{t}(\omega)$ where $\xi^{o}$ is the optional projection of $\xi$, $\xi^{o}_{t}=\Esp{\xi_{t}|\Fc_{t}}$ for $t\le T$. Set $\mu_{A}(B)=\mu(B\times A)$ for $B\in \Bc([0,T])$. Then, $\mu_{A}$ is an optional random measure on $[0,T]$ by our assumption on $\mu$. Moreover, $\mu_{A}(Y)=\mu(Y\1_{A})$ for $Y=\xi,\xi^{o}$. 
It then follows from  \cite[Chapter VI.2]{DellacherieMeyer} that \reff{ProjectOpt-Equal}  holds. The monotone class theorem allows to conclude in the case where $X$ is just measurable and bounded. The general case is obtained by a standard truncation argument.  
\ep
\begin{Proposition} \label{prop : joint measurablity}
Let $E$ be a compact metrizable topological space  and $\Gc$ a sub $\sigma$-algebra of $\Fc$. 
 \begin{enumerate}[(a.)]
\item The following bi-linear form is continuous
$$M_{+\sigma}(E) \times C_{\beta}(E) \ni (\nu,g) \mapsto \nu(g) \in \R.$$
 \item Fix  $g\in L^{0}(\Gc; C_{\sigma}(E))$  and  $\mu\in L^{0}(\Gc;M_{+\sigma}(E))$. Then  $\mu(g) \in L^{0}(\Gc)$.
 \item Fix $\mu \in L^{0}(\Gc; M_{+\sigma}(E))$. Then,  the map $(\omega,g)\in \Omega\times C(E)\mapsto \mu(\omega)(g)$ is $\Gc\otimes \Bc(C_{\sigma}(E))/\Bc(\R)$-measurable.
 \end{enumerate}
\end{Proposition}
\textbf{Proof:}
(a.) By Pettis' theorem,  weakly-measurable and strongly measurable $C(E)$-valued random variables coincides. We can then assume that $g$ is strongly measurable. Let  $(h_n)_{n}$ (resp. $(\mu_n)_{n}$)  be a convergent sequence in the Banach space $C_{\beta}(E)$  (resp. the Polish space  $M_{+\sigma}(E)$ (see Corollary \ref{cor: M+ polish})) converging to $h$ (resp.  $\mu$). The triangular inequality implies that  $|\mu_n(h_m)-\mu(h)|\leq |(\mu_n-\mu)(h)| +|\mu_n(h_m-h)|$ for all $n,m\ge 1$. The first term on the r.h.s. converges to $0$ by weak* continuity. The second converges to $0$ by norm convergence in $C(E)$ and norm boundedness of $(\mu_n)_{n\ge 1}$ (since weak*-convergent). This proves the continuity of  the bi-linear form.

(b.) This assertion now follows by continuous composition of measurable mappings.

(c.) Also here the continuity of the bi-linear form and the composition with a measurable mapping gives the result.
\ep

\subsection{Some topological properties of the solvency cones}

We now establish some topological properties of the solvency cones. Many arguments below are   inspired by standard texts, see e.g.~\cite{Bour INT ch 1-4}. Since a deterministic set-up is sufficient here, we only consider deterministic transaction costs $\lambda$, but we consider a slightly more general context in terms of spaces than in the preceding sections.
Namely, we consider two spaces $X$ and $Y$ satisfying
\be\label{eq: space X Y}
\mbox{ $X$ is a compact metrizable space    and   $Y:=\T \times X$}
\ee
where $\T=[0,T]$ for some $T \in [0,\infty)$. For $\lambda \in C_{+}(\T\x X^2)$ the cone $K(\lambda)$ is now defined (cf. Sec.\ref{sec: olvency cones}) to be the closure in $M_\sigma(Y)$ of the cone
 \begin{equation} \label{eq: solv cone in  M(K) general 1}
\mathrm{cone}\{(1+\lambda_t(x,y))  \delta_t\otimes \delta_x -  \delta_t \otimes \delta_y, \,  \delta_t \otimes \delta_x :    (t,x,y) \in \T\x X^2\}.
\end{equation}
The dual cone $K'(\lambda) \subset C(Y)$ of the cone $K(\lambda)$ in  $M_\sigma(Y)$ is
 \begin{equation} \label{eq: dual cone in  C(K) general} 
K'(\lambda)
 =\{f \in C_{+}(Y) :  f(t,y) \leq (1+\lambda_t(x,y)) f(t,x), \; \forall (t,x,y) \in \T\x X^2\}.
 \end{equation}
We note that $K(\lambda) \subset M(Y)$ is the dual cone of the cone $K'(\lambda)$ in  $C_\sigma(Y)$ and also of the cone $K'(\lambda)$ in  $C_\beta(Y).$
Let us define
\be\label{eq: def lambda int} 
\Lambda_{\mathrm{int}}:=\{\lambda \in C_{+}(\T\x X^2) \mbox{ s.t. } \interior{(\tdual{K}(\lambda))} \neq \emptyset\},
\ee
in which the interior is taken in $C_\beta(Y)$.

\begin{Remark}\label{rem: cond necessaire int non vide} For later use, note that   {$\lambda_{t}(x,y)\ge 0$ and
$\interior{(\tdual{K}(\lambda))} \neq \emptyset$ imply that}    $(1+\lambda_{t}(x,y))^{-1}<1+\lambda_{t}(y,x)$ or equivalently $\lambda_{t}(x,y)+\lambda_{t}(y,x)>0$, for $(t,x,y)\in \T\x X^2$, see \reff{eq: int K'}.
{An easy consequence is   that}  the cone in \reff{eq: solv cone in  M(K) general 1} coincides with 
 \b*
\mathrm{cone}\{(1+\lambda_t(x,y))  \delta_t\otimes \delta_x -  \delta_t \otimes \delta_y  :    (t,x,y) \in \T\x X^2\}
\e*
whenever $\interior{(\tdual{K}(\lambda))} \neq \emptyset$. 
\end{Remark}
  
\begin{Lemma} \label{lm: completeness of K}
Fix $\lambda \in \Lambda_{\mathrm{int}}$. In the space $M(Y)$, the cone $K(\lambda)$ is complete for the uniform structure deduced from the weak* topology.
\end{Lemma}

\proof 1. Linear functionals on  $C(Y)$, positive w.r.t.  the order defined by the cone $K'(\lambda)$ are strongly continuous. More precisely, letting $C^*(Y)$ denote the algebraic dual of $C(Y)$, we have 
\begin{equation} \label{eq: positive mu}
 \text{if } \mu \in C^*(Y) \text{ and }  \mu(f) \geq 0\; \forall \, f  \in K'(\lambda) \; \text{ then } \mu \in K(\lambda).
\end{equation}
This is seen as follows: By hypothesis  $K'(\lambda)$ has non-empty interior in $C_\beta(Y)$, so $\mu \in M(Y)$, the topological dual of $C_\beta(Y)$ (cf. \cite[Theorem in Ch. V Sec. 5.5]{Schaefer}). Statement  \reff{eq: positive mu} now follows, since the dual cone of the cone $K'(\lambda)$ in  $C_\beta(Y)$ is $K(\lambda).$

2. Let $\Uc$ be a Cauchy filter for the weak* uniform structure  on $K(\lambda)$. Then for all $f \in C(Y)$ the limit $\nu(f):=\lim_{\mu, \Uc}\mu(f)$ exists, so $\nu \in C^*(Y)$. Moreover  $\nu(f) \geq 0$ if $f \in K(\lambda)$, which together with 
\reff{eq: positive mu} shows that $\nu \in K(\lambda)$. \qed

\begin{Lemma} \label{lm: equal topologies on K}
Let $\lambda \in \Lambda_{\mathrm{int}}$ and let $V$ be a dense subspace of $C_\beta(Y)$. The topologies on $K(\lambda)$ induced by  $\sigma(M(Y),V)$ and $\sigma(M(Y),C(Y))$ are identical.
\end{Lemma}
\textbf{Proof:} Let $\Uc$ be an ultra filter on $K(\lambda)$, converging to a measure $\nu$ for the topology $\sigma(M(Y),V)$. Since the topology $\sigma(M(Y),C(Y))$ is finer than  $\sigma(M(Y),V)$, it is enough to prove that $\Uc$ converges to $\nu$ also in 
$\sigma(M(Y),C(Y))$.

Since the cone $K'(\lambda)$ has an interior point $u_1$, there exists $r_1>0$ such that $u_1+r_1B(0,1) \subset \mathrm{int}(K'(\lambda))$, where $B(0,1)$ is the open unit ball of $C_\beta(Y)$ centered at $0$. As $V$ is dense in $C_\beta(Y)$, we can then choose $u \in V \cap (u_1+r_1B(0,1))$ and $r >0$ such that $u+rB(0,1) \subset \mathrm{int}(K'(\lambda))$.

Then,  $\mu(u+rg) \geq 0$  for all $\mu \in K(\lambda)$ and $g \in B(0,1)$, which leads to $|\mu(g)| \leq \mu(u)/r$.
As $u \in V$, it follows from the definition of the topology $\sigma(M(Y),V)$ that there exists a set $N \in \Uc$ such that
$$
\; 0 \leq \mu(u) \leq \nu(u) +1 \; \;\forall \mu \in N.
$$
According to the last two inequalities, $\sup_{\mu \in N}|\mu(g)| \leq (\nu(u)+1)/r$ for all $g \in B(0,1)$, which shows that $N$ is weak*-bounded.

The topologies on $N$ induced by  $\sigma(M(Y),V)$ and $\sigma(M(Y),C(Y))$ are identical, since   $N$ is weak*-bounded (cf. \cite[Proposition 17, Ch. III, \S 1, nr.  10]{Bour INT ch 1-4}. The ultrafilter $\Uc_N$ on $N$ induced by $\Uc$  converges to $\nu$ in the topology induced by $\sigma(M(Y),V)$, so it also converges to $\nu$ in the topology induced by $\sigma(M(Y),C(Y))$. \qed
\vs5

The cone $K(\lambda) \subset M(Y)$  endowed with the induced topology as a subspace of $M_\sigma(Y)$  is denoted $K_\sigma(\lambda)$ from now on.
\begin{Proposition} \label{prop: K is polish}
 If $\lambda \in \Lambda_{\mathrm{int}}$ then $K_\sigma(\lambda)$ is a Polish space.
\end{Proposition}
\textbf{Proof:}  The topological space $Y$ being compact and metrizable, it is separable (cf. \cite[Propositions 12 and 16, Ch. IX, \S2]{Bourb TG 5-10}). Let the countable set $D=\{y_n \in Y \,:\, n \in \N^*\}$ be dense in $Y$ and define the set of measures
$$
A_1=\{\mu \in M(Y) \,:\, \mu=\sum_{i \in I}a_i \delta_{y_i}, \; y_i \in D, \; a_i \in \Q \text{ and } I \text{ finite} \}.
$$
$A_1$ is dense in $M_\sigma(Y)$. It follows directly from the definition of $K(\lambda)$ and \reff{eq: solv cone in  M(K) general 1} that the  set $A=A_1 \cap K_\sigma(\lambda)$ is dense in $K_\sigma(\lambda)$. The topological space $K_\sigma(\lambda)$ is therefore separable.

The space $C_\beta(Y)$ is separable, since  $Y$ is compact and metrizable (cf. \cite[Theorem 1, Ch. X, \S3]{Bourb TG 5-10}). Let $\tilde C$ be a countable and dense subset of $C_\beta(Y)$ and let $V$ be the linear hull of  $\tilde C$.

According to Lemma \ref{lm: equal topologies on K}, the topologies on  $K_\sigma(\lambda)$ induced by $\sigma(M(Y),V)$ and $\sigma(M(Y),C(Y))$ are identical. Since the (algebraic) dimension of $V$ is countable, it follows that every measure $\mu \in K_\sigma(\lambda)$ has a countable local base of open neighbourhoods
$$ \Bc(\mu) = \{B_n(\mu) \,:\, n \in N^*\}.$$
We choose a sequence $(\mu_n)_{n \geq 0}$ in $M(Y)$ such that $A=\{\mu_n \,:\, n \in N^*\}.$ The family of open sets
$$
\Bc = \{B_n(\mu_m) \,:\, n,m \in N^*\}
$$
is then a countable base of the  topology of $K_\sigma(\lambda).$ Since $K_\sigma(\lambda)$ is locally compact, it now follows that it is metrizable (cf. \cite[Corollaire, Ch. IX, \S2, nr. 9]{Bourb TG 5-10}).
Finally, $K_\sigma(\lambda)$ is complete according to Lemma \ref{lm: completeness of K}.
 \qed

\vs5

Since $M_{+\sigma}(Y)$ is a closed subset of $K_\sigma(\lambda)$, when $\lambda \in \Lambda_{\mathrm{int}}$, the following is deduced from the   above by  setting $T=0$.
\begin{Corollary} \label{cor: M+ polish}
$M_{+\sigma}(X)$ is a Polish space.
\end{Corollary}

 \subsection{A measurable selection result for trading strategies}

We now establish a measurable selection result. It is used in the proof of Proposition \ref{prop : PropFondSurX V2} to establish that simple strategies are admissible. 

This requires the   introduction of  some additional notations and of an elementary notion of deterministic causality described by progressive measurability, but without reference to the filtered probability space $(\Omega,\Fc,\F,\P)$.  

As in the preceding section, $X$ and $Y$ are given as in \reff{eq: space X Y},  while $\Lambda_{\mathrm{int}}$ is defined in \reff{eq: def lambda int}.

Let $\Lambda_{\mathrm{int},\beta}$  be $\Lambda_{\mathrm{int}}$ endowed with the induced topology as a subspace of  $C_{\beta}(\T\x X^2)$. In all this section, we fix $$\hat \lambda \in \Lambda_{\mathrm{int}},$$   and define $\hat \Lambda$ (resp. $\hat \Lambda_\beta$) as the subset of $\Lambda_{\mathrm{int}}$ (resp. subspace of $\Lambda_{\mathrm{int},\beta}$) of elements $\lambda \in \Lambda_{\mathrm{int}}$ such that $\lambda \geq \hat\lambda.$

The topological space $$A:=\hat\Lambda_\beta \times M_{+ \sigma}(\T\x X^2)$$  is Polish since this is the case of $ M_{+ \sigma}(\T\x X^2)$ (apply Corollary \ref{cor: M+ polish}).

Using that $K(\lambda) \subset K(\hat\lambda)$ for $\lambda \in \hat \Lambda$, we define the subspace  $B \subset \hat\Lambda_\beta \times K_\sigma(\hat\lambda)$ by
\be\label{eq: def B}
B =\bigcup_{\lambda \in \hat\Lambda} \{\lambda\} \times  K(\lambda).
\ee
Let $\hat\rho$ be a metric for the Polish space $K_\sigma(\hat\lambda)$, see Proposition \ref{prop: K is polish}. Since $K(\lambda)$ is a weak*-closed subspace of $K(\hat\lambda)$, it is also a complete metric space for $\hat\rho.$ Let us define $\rho^B(\lambda_1,\mu_1,\lambda_2,\mu_2)=\|\lambda_1-\lambda_2\|_{C(\T\x X^2)} +\hat\rho(\mu_1,\mu_2)$ for $(\lambda_{i},\mu_{i})\in C(\T\x X^2)\x M_{+ \sigma}(Y)$, $i=1,2$.  Then,  $B$ is a complete metric space for the metric $\rho^B$.

Let $H_\lambda$ be defined by \reff{eq: def H} for   a given $\lambda$:
\be\label{eq: def Hlambda}
H_{\lambda}(f)(s,x,y):= f(s,y)-  f(s,x)(1+\lambda_s(x,y)),\; (s,x,y) \in \T\x\bRpd.
\ee
 We can now define the mapping $I:A \rightarrow B$ by
\be\label{eq: def I}
I(\lambda,L)=(\lambda, -L \circ H_\lambda).
\ee
 We recall that, given   two locally compact Hausdorff spaces $U$ and  $V$,  a mapping of $U$ into $V$ is called proper when it is continuous and the inverse image of every compact set is compact.
\begin{Lemma} \label{lm: mapping I}
 The mapping $I:A \rightarrow B$ is proper, closed and surjective.
\end{Lemma}
\textbf{Proof:}
1.  We first show the continuity. The first component of $I$ is the identity mapping on $\hat \Lambda$, so it is continuous. For all $f \in C(\T \x X)$, the mapping $\hat \Lambda_\beta \ni  \lambda \mapsto H_\lambda(f) \in C_{\beta}(\T \x X^2)$ is continuous. The continuity of the second component of $I$ now follows from (a.) of Proposition \ref{prop : joint measurablity}.

2.  We now show that $I$ is proper.  Suppose that $E^B$ is a compact subset of $B$. It is enough to prove that $E^A:=I^{-1}(E^B)$ is compact. This is true if $E^A$ is empty. Suppose that $E^A$ is not empty. To prove that $E^A$ is compact it is enough to establish that every sequence $(a_n)_{n\geq 1}$ in  $E^A$ has a convergent sub sequence with limit $a \in E^A$. For $a_n=(\lambda_n,L_n)$, we set $b_n=I(a_n)=(\lambda_n,\mu_n)$. Since $E^B$ is compact, the sequence  $(b_n)_{n\geq 1}$ in  $E^B$ has a convergent sub sequence, which after re-indexing we also denote by $(b_n)_{n\geq 1}$, with limit $b \in E^B$. Then, after re-indexing the corresponding sub sequence of $(a_n)_{n\geq 1}$, $b_n=I(a_n)$ and $b=\lim_nI(a_n)$. As to be established below, for fixed $\hat f \in \mathrm{int}(K'(\hat \lambda))$ there exists a constant $C>0$  such that for all $(\lambda, L) \in A$ and $(\lambda, \mu)=I(\lambda, L)$
\begin{equation} \label{eq: L bound}
 \|L\|_{M(\T\x X^{2})} \leq C \mu(\hat f).
\end{equation}
The sequence $(\mu_n)_{n \geq 1}$ is weak*-bounded, so $\mu_n(\hat f) \leq c$  for some $c \geq 0$.  The sequence $(a_n)_{n\geq 1}$
then satisfies $ \|L_n\|_{M(\T\x X^{2})} \leq c \, C$, and by weak*-compactness it therefore exists a sub sequence, also called  $(a_n)_{n\geq 1}$ after re-indexing, in $E^A$ converging to a limit $a \in A$. By the continuity of $I$, it follows that $I(a)=b \in E^B$.  So  the original sequence $(a_n)_{n\geq 1}$ has a sub-sequence converging to  $a \in E^A$.

It remains to prove \reff{eq: L bound}. For   $\hat f \in \mathrm{int}(K'(\hat \lambda))$ and $(\lambda, L) \in A$, so that $\lambda\ge \hat \lambda$ in particular, we have      $ -H_\lambda(\hat f)  \geq -H_{\hat\lambda}(\hat f) \geq \epsilon$ for some constant $\epsilon>0$, recall  \reff{eq: int K'} and \reff{eq: def H}. This gives
$$
\mu(\hat f) = L(-H_\lambda(\hat f)) \geq L(-H_{\hat\lambda}(\hat f)) \geq L(\epsilon) = \epsilon  \|L\|_{M(\T\x X^{2})},
$$
which proves \reff{eq: L bound} with $C=1/\epsilon$.

3. $I$ is closed since it is  proper and $A$ and $B$ are locally compact.

4.  We finally show that $I$ surjective. It is enough to prove that, for given $\lambda \in \hat \Lambda$,  the function $M_+(\T\x X^2) \ni \lambda \mapsto -L  \circ H_\lambda$ is onto $K(\lambda)$. According  to its definition through \reff{eq: solv cone in  M(K) general 1}, Remark \ref{rem: cond necessaire int non vide} and the definition of  $\hat \Lambda$, $K(\lambda)$   is the closure in  $M_\sigma(\T \x X^{2})$ of the cone
 $$
\mathrm{cone}\{ -(\delta_t\otimes \delta_x \otimes \delta_y)  \circ H_\lambda :    (t,x,y) \in \T\x X^2\},
$$ 
recall the definition of $H_{\lambda}$ in \reff{eq: def Hlambda}. This shows that $K(\lambda)$ is the $M_\sigma(\T \x X^{2})$ closure of $\check{K}(\lambda):= \{ -L  \circ H_\lambda : L \in M_+(\T\x X^2)\}$.
 We shall in fact show that 
 \begin{equation} \label{eq: solv cone in  M(K) general 2}
K(\lambda)=\check{K}(\lambda)=\{ -L  \circ H_\lambda :     L \in M_+(\T\x X^2)\}.
\end{equation}
Let the sequence $(\mu_n)_{n \geq 1}$ in $\check{K}(\lambda)$ converge to $\mu$ in the Polish space $K_\sigma(\lambda)$, recall Proposition \ref{prop: K is polish}. With $L_n$ such that $\mu_n=-L_n  \circ H_\lambda$, inequality (\ref{eq: L bound}) and the  weak*-boundedness of $(\mu_n)_{n \geq 1}$ show that the exists   $m>0$ for which  $\|L_n\|_{M(\T\x X^2)} \leq m$ for all $n$. By weak*-compactness, there is then a sub-sequence, also denoted  $(L_n)_{n \geq 1}$ after re-indexing, converging to some $L \in M_+(\T\x X^2)$. By weak*-continuity, $\mu = -L  \circ H_\lambda$. This proves \reff{eq: solv cone in  M(K) general 2}. 
\ep
\\

In order to introduce a progressive $\sigma$-algebra on $\T \x A$, let $C^{\text{Pr}}(\T \x A)$ be the subset of functions $f \in C(\T \x A)$ such that $f(t,a)=f(t,a')$ if $a(s)=a'(s)$ for all $s\in [0,t]$, for $t\in \T$. Let $C^{\text{Pr}}(\T \x B)$ be defined similarly with $B$ in place of $A$.  The topological space $A^{\text{Pr}}$ (resp.  $B^{\text{Pr}}$)  is $\T \x A$ (resp.  $\T \x B$) endowed with the coarsest topology for which all functions in $C^{\text{Pr}}(\T \x A)$ (resp. $C^{\text{Pr}}(\T \x B)$) are continuous. 
 The mapping $I^{\text{Pr}} :  A^{\text{Pr}} \rightarrow  B^{\text{Pr}}$ is defined by
\begin{equation} \label{eq : I Prog}
 I^{\text{Pr}}(t,\lambda, L)=(t, I(\lambda,L)),
\end{equation}
where $I$ is defined in \reff{eq: def I}.

 For $t \in \T$ consider the canonical projection
$$
A \ni (\lambda,L) \mapsto(\lambda,L)|_{[0,t] \x X^2} \in C_\beta ([0,t]\x X^2)\times M_{+ \sigma}([0,t]\x X^2).
$$
For  $t \in \T$,  $\Fc^A_t$ is the inverse image of the Borel  $\sigma$-algebra $\Bc(A)$ under this projection and $\F^A:=(\Fc^A_t)_{t \in \T}$ defines a filtration of $A$ (when endowed with its conventional Borel measurable space structure). 
Similarly,  the $\sigma$-algebra $\Fc^B_t$, for $t \in \T$, is the inverse image of $\Bc(B)$ under the  the canonical projection
\be\label{eq: def cano proj}
B \ni (\lambda,\mu) \mapsto(\lambda,\mu)|_{[0,t] \x X} \in C_\beta ([0,t]\x X)\times (M_{\sigma}([0,t]\x X) \cap  K_\sigma(\hat\lambda)).
\ee
$\F^B:=(\Fc^B_t)_{t \in \T}$ defines a filtration of $B$.

We note that the spaces  $A^{\text{Pr}}$ and   $B^{\text{Pr}}$ are in general not Hausdorff, since in general $C^{\text{Pr}}(\T \x A)$ and $C^{\text{Pr}}(\T \x B)$ do not separate points in $\T \x A$ and  $\T \x B$ respectively.  For this reason, we shall need to use a suitable notion of equivalent classes. To define them, we first introduce the map $i : \T \x \hat \Lambda_\beta \rightarrow \hat \Lambda_\beta$ defined by
$$
i_t(\lambda)(x,y)=
\begin{cases}
 \lambda_s(x,y) \text{ if } s \in [0,t] \\  \max(\hat\lambda_s(x,y),\lambda_t(x,y))  \text{ if } s \in (t,T],
\end{cases}
$$
for all $(t,x,y) \in \T \x X^2$. 

We then define sets of progressive processes $\tilde{A}^{\text{Pr}}$ and $\tilde{B}^{\text{Pr}}$, representing the equivalence classes, and a  mapping $\tilde{I}^{\text{Pr}}:\tilde{A}^{\text{Pr}} \rightarrow \tilde{B}^{\text{Pr}}$ by
\be  \label{eq : tilde A Prog}
&\tilde{A}^{\text{Pr}}= \{(t,i_t(\lambda), L_t) \in \T \x A : (\lambda,L) \in A \}, \text{ where } L_t= L|_{[0,t] \x X^2},&
\\
 \label{eq :  tildeB Prog}
&\tilde{B}^{\text{Pr}}= \{ (t,i_t(\lambda), \mu_t) \in \T \x B : (\lambda,\mu) \in B \}, \text{ where } \mu_t= \mu|_{[0,t] \x X},&
\\ 
 \label{eq : tilde I Prog}
 & \tilde{I}^{\text{Pr}}~:~(t,\alpha,N)\in\tilde{A}^{\text{Pr}}\mapsto (t,\alpha, -N \circ H_{\alpha})\in \tilde{B}^{\text{Pr}}.&
\ee
\begin{Theorem}\label{th: measurable selection L}
The mapping  $\tilde{I}^{\text{Pr}}:\tilde{A}^{\text{Pr}} \rightarrow \tilde{B}^{\text{Pr}}$ is proper, closed and surjective and it has a Borel measurable  right inverse  $\tilde{J}^{\text{Pr}} :\tilde{B} ^{\text{Pr}} \rightarrow \tilde{A}^{\text{Pr}}$.
\end{Theorem}
\textbf{Proof:}
1. The continuity follows directly from the continuity of $I$, see Lemma \ref{lm: mapping I}.

2. We now show that $\tilde{I}^{\text{Pr}}$ is proper. Let $C$ be a compact subset of $\tilde{B}^{\text{Pr}}$ and let $(t^n,\alpha^n, N^n)_{n\geq 1}$ be a sequence in $(\tilde{I}^{\text{Pr}})^{-1}(C)$.  By compactness, the sequence $(t^{n},\alpha^{n},\nu^{n})_{n\ge 1}$ $= $ $(\tilde{I}^{\text{Pr}}(t^n,\alpha^n,N^n))_{n\geq 1}$ in $C$ has a convergent sub-sequence with a limit $(t,\alpha,\nu) \in C$ and, possibly after extracting a sub-sequence, we can suppose that  $(t^{n},\alpha^{n},\nu^{n})_{n\ge 1}$ converges to $(t,\alpha,\nu)$.

The set $C_1:=\{(\alpha, \nu),(\alpha^n, \nu^n) : n\geq 1\}$ is a compact subset of $B$, so it follows from Lemma \ref{lm: mapping I} that its inverse image under $I$ is compact.  Hence, after possibly extracting  a convergent sub-sequence, we can suppose that $(\alpha^n, N^n)_{n\geq 1}$ converges to some $(\alpha, N)$ in $I^{-1}(C_1)$, so that the sequence $(t^n,\alpha^n, N^n)_{n\geq 1}$ converges to $(t,\alpha, N) \in \T \x A$. Since $i_{t^n}(\alpha^n)=\alpha^n$ and $\text{supp}(N^n) \subset [0,t_{n}] \x X^2$, it follows by continuity that  $i_t(\alpha)=\alpha$ and  $\text{supp}(N) \subset [0,t] \x X^2$, which proves that $(t,\alpha, N) \in \tilde{A}^{\text{Pr}}$.

3. $\tilde{I}^{\text{Pr}}$ is closed since it is proper and $\tilde{A}^{\text{Pr}}$ and $\tilde{B}^{\text{Pr}}$ are locally compact.

4.   $\tilde{I}^{\text{Pr}}$ is surjective. To see this, fix  $(t,\alpha,\mu) \in \tilde{B}^{\text{Pr}}$. Since $I: A \rightarrow B$ is surjective (Lemma \ref{lm: mapping I}), there exists $(\alpha,L) \in A$ such that $I(\alpha,L)=(\alpha,\mu)$.  Since $(t,\alpha,\mu) \in \tilde{B}^{\text{Pr}}$, we must have $ (t,\alpha,L) \in \tilde{A}^{\text{Pr}}$.   Then $\tilde{I}^{\text{Pr}}(t,\alpha, L)=(t,\alpha, \mu)$, by (\ref{eq : tilde I Prog}).

5.  Since  $\tilde{I}^{\text{Pr}}$ is  closed and surjective,  the inverse image $(\tilde{I}^{\text{Pr}})^{-1}$ defines an upper hemicontinuous correspondence $\varphi$, i.e. a function of $\tilde{B}^{\text{Pr}}$ into the set of subsets of $\tilde{A}^{\text{Pr}}$, cf. \cite[Theorem 17.7]{Aliprantis-Border}. Its upper inverse  $\varphi^u : \tilde{A}^{\text{Pr}} \rightarrow 2^{\tilde{B}^{\text{Pr}}}$ is explicitly given by $\varphi^u(x)=\{\tilde{I}^{\text{Pr}}(x)\}$. From the closeness of $\tilde{I}^{\text{Pr}}$ it now follows that $\varphi$ is weakly measurable correspondence (see \cite[Definition 18.1  and the discussion below]{Aliprantis-Border}). Then $\varphi$ has a measurable selector  $\tilde{J}^{\text{Pr}}$,  according to the selection theorem  \cite[Theorem 18.13]{Aliprantis-Border}.  \qed\\

Let  $\hat J: \tilde B^{\text{Pr}} \rightarrow  M_{+ \sigma}(\T\x X^2)$ be the third component of  $\tilde{J}^{\text{Pr}}$, i.e. $\tilde{J}^{\text{Pr}}(t,\alpha,\nu)$ $=(t,\alpha, \hat J (t,\alpha,\nu))$ for all $(t,\alpha,\nu) \in  \tilde B^{\text{Pr}}$.  Due to the definition of   $\tilde{A}^{\text{Pr}}$,   $\tilde{B}^{\text{Pr}}$ and $\tilde{I}^{\text{Pr}}$, it follows that, for all $(t, \lambda,\mu) \in \T \x B$,
\begin{equation} \label{eq: hat J}
 \hat J (t,i_t(\lambda),\mu |_{[0,t] \x X})= \hat J (T,\lambda,\mu) |_{[0,t] \x X^2} .
\end{equation}
The left hand side of this formula defines a $M_{+ \sigma}(\T\x X^2)$ valued progressive process w.r.t. the filtration $\F^B$ of $B$, which is then also the case for the right hand side. 

We now define the Borel measurable function 
\be\label{eq: def J}
J : (\lambda, \mu)\in B \rightarrow \hat J (T,\lambda,\mu) \in M_{+ \sigma}(\T \x X^2). 
\ee
 One can then sum up the above as follows. 
\begin{Corollary}\label{cor: measurable selection L} Let $J$ and $B$ be defined as in \reff{eq: def J} and \reff{eq: def B}.
 
The  process $\T \x B \ni (t, \lambda, \mu) \mapsto J(\lambda, \mu)|_{[0,t] \x X^2} \in M_{+ \sigma}(\T \x X^2)$ is progressively measurable w.r.t. the filtration  $\F^B=(\Fc_{t}^{B})_{t\in \T}$, in which   $\Fc_{t}^{B}$ is defined as   the inverse image of $\Bc(B)$ under   the canonical projection \reff{eq: def cano proj}.
\end{Corollary}

\end{document}